\pdfoutput=1  
\documentclass[]{article}  
\usepackage{url,float}
\usepackage{graphicx}
\usepackage{amsmath}
\usepackage{amsfonts}
\usepackage{amssymb}
\usepackage{latexsym}

\newcommand{\hide}[1]{}

\newcommand{\ABox}{
\raisebox{3pt}{\framebox[6pt]{\rule{6pt}{0pt}}}
}
\newenvironment{proof}{{\bf Proof:}}{\hfill\ABox}

\newtheorem{theorem}{{\bf Proposition}}

\newtheorem{lemma}{Lemma}
\newtheorem{conjecture}[theorem]{Conjecture}

\newcommand{\lemlab}[1]{\label{lemma:#1}}
\newcommand{\thmlab}[1]{\label{thm:#1}}

\newcommand{\figlab}[1]{\label{fig:#1}}
\newcommand{\seclab}[1]{\label{sec:#1}}
\newcommand{\secref}[1]{\ref{sec:#1}}
\newcommand{\figref}[1]{\ref{fig:#1}}

\def\a{{\alpha}}
\def\b{{\beta}}
\def\q{{\theta}}
\def\r{{\rho}}
\def\s{{\sigma}}
\def\l{{\lambda}}
\def\o{{\omega}}
\def\R{{\mathbb{R}}}

\newcommand{\squeezelist}{\setlength{\itemsep}{0pt}}

\title{%
Spiral Unfoldings of Convex Polyhedra
} 

\author{%
Joseph O'Rourke%
    \thanks{Departments of Computer Science, and Mathematics, Smith College, Northampton, MA
      01063, USA.
      \protect\url{orourke@cs.smith.edu}.}
}

\begin{document}
\maketitle

\begin{abstract}
The notion of a spiral unfolding of a convex polyhedron, a special type of
Hamiltonian cut-path, is explored.
The Platonic and Archimedian solids all have nonoverlapping
spiral unfoldings, although overlap is more the rule than the exception
among generic polyhedra. The structure of spiral unfoldings is described,
primarily through analyzing one particular class, the polyhedra of revolution.
\end{abstract}

\section{Introduction}
\seclab{Introduction}
I define 
a \emph{spiral} $\s$ on the surface of a convex polyhedron $P$ as a
simple (non-self-intersecting) polygonal path 
$\s=(p_1,p_2,\ldots,p_m)$
which includes every vertex $v_j$ of
$P$ (so it is a Hamiltonian path), and so when cut permits
the surface of $P$ to be unfolded flat into $\R^2$.
(Other requirements defining a spiral will be discussed below.)
The starting point for this investigation was
Figure~\figref{CubeLay},
an unfolding of a spiral cut-path on a tilted cube,
Figure~\figref{Cube3d}.
\begin{figure}[htbp]
\centering
\begin{minipage}{.48\textwidth}
\centering
\includegraphics[width=0.9\linewidth]{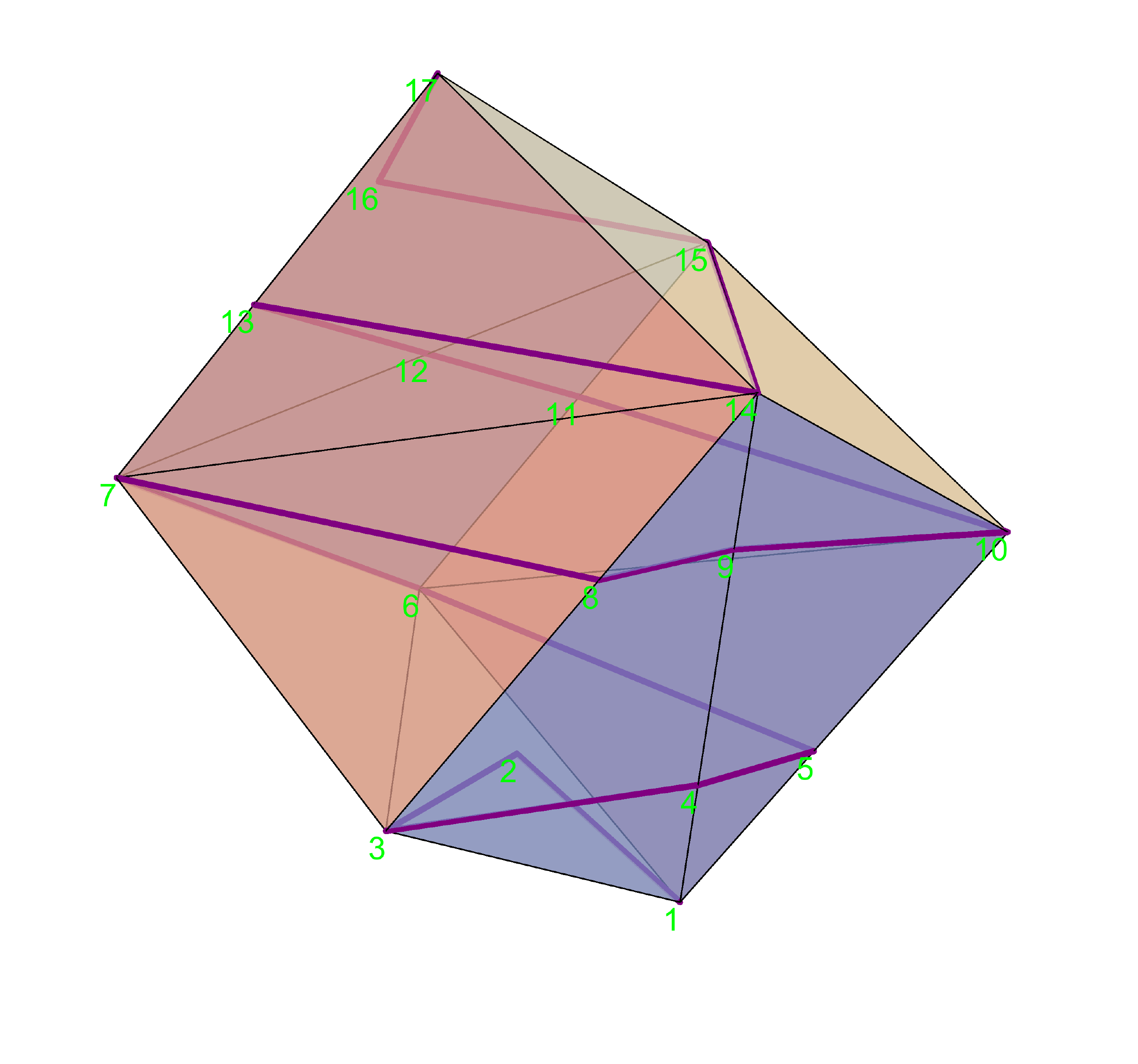}
\caption{Cube with spiral cut-path $\s$.
}
\figlab{Cube3d}
\end{minipage}%
\quad%
\begin{minipage}{.48\textwidth}
\centering
\includegraphics[width=1.0\linewidth]{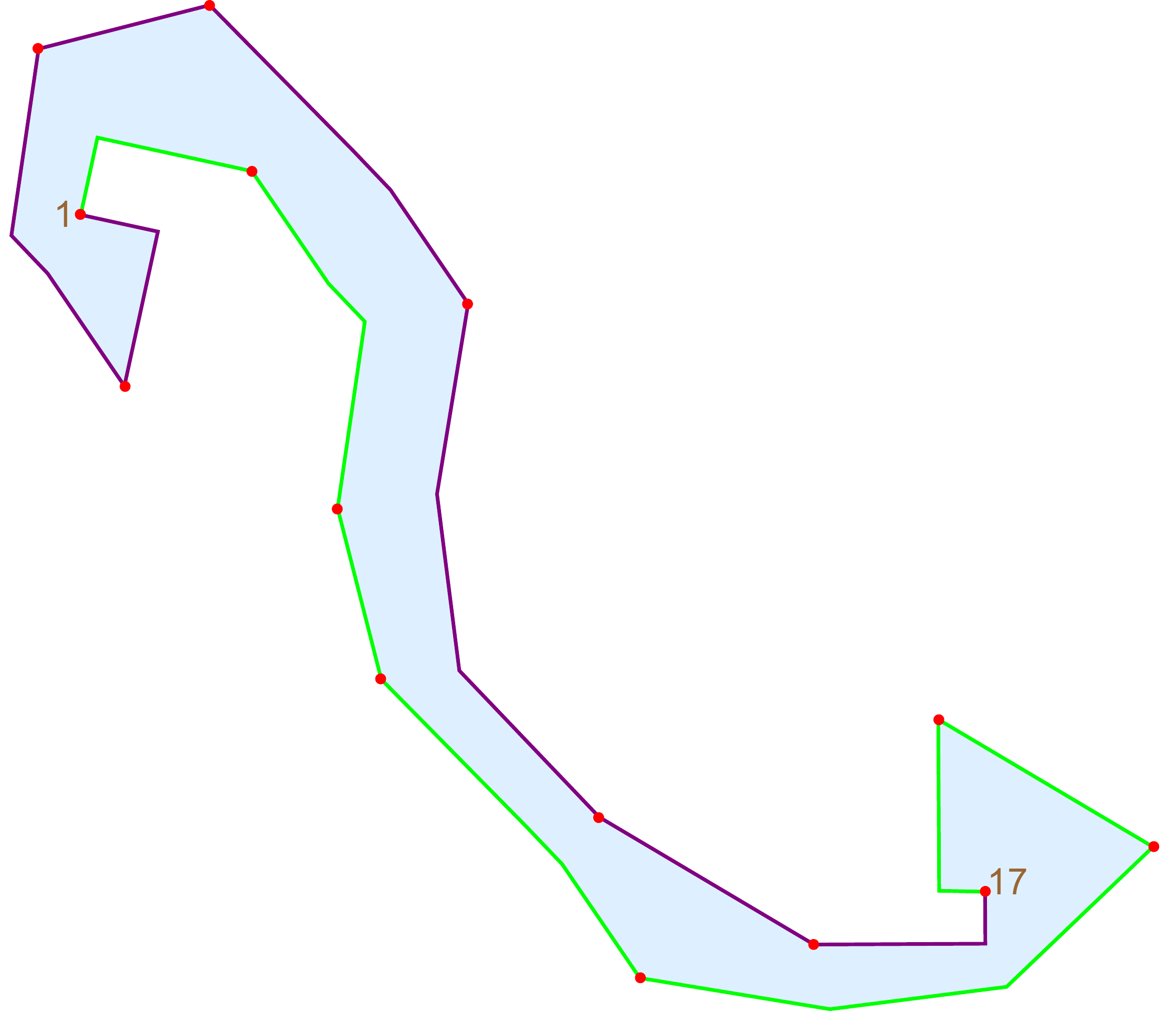}
\caption{Spiral unfolding of cube. Vertices are marked with red dots.
$\r$ is purple; $\l$ is green.
}
\figlab{CubeLay}
\end{minipage}
\end{figure}
The cut-path $\s$ on $P$ unfolds to two paths $\r$ and $\l$ in the plane,
with the surface to the right of $\r$ and to the left of $\l$.
Folding the planar layout by joining (``gluing") $\r$ to $\l$
along their equal lengths results in the cube,
uniquely results by Alexandrov's theorem.
Note that the external angle at the bottommost vertex (marked $1$ in Figure~\figref{CubeLay})
and the topmost vertex (marked $17$, because $\s$ has 
$16$ segments and $m=17$
corners\footnote{We reserve the term ``vertices" to refer to $P$'s vertices,
and use ``corners" for the turns of $\s$.})
is $90^\circ$, which is the Gaussian curvature at those cube vertices.
Note also that there are $8$ vertices of $P$ along both $\r$ and $\l$, with one shared
at either end.

The notion of restricting unfoldings of convex polyhedra
by following a Hamiltonian path along polyhedron edges was introduced by 
Shephard 40 years ago~\cite{s-cpcn-75}.
Shepard noted that not every polyhedron has such a Hamiltonian edge-unfolding,
because not every polyhedron $1$-skeleton has a Hamiltonian path
(e.g., the rhombic dodecahedron does not have such a path).
Here we are not restricting cuts to polyhedron edges.
A single Hamiltonian cut-path, not necessarily following polyhedron edges,
leads to what was memorably christened as
a \emph{zipper unfolding} in~\cite{lddss-zupc-10}.
Those authors posed the still-open problem of whether or not
every convex polyhedron has a nonoverlapping zipper unfolding.
The investigation reported here does not make an advance on 
this question, as spiral unfoldings are a special subclass of zipper unfoldings,
and nonoverlap is rare (as we will see in Section~\secref{Overlapping}).
Instead we pursue these unfoldings for their intrinsic,
almost aesthetic, interest.
In keeping with this attitude, we reach no grand conclusions, and do not offer formal proofs
of claims.
For terminology and background (e.g., Alexandrov's theorem, Gaussian curvature), see~\cite{do-gfalop-07}.

The cube unfolding in Figure~\figref{CubeLay} suggests seeking spiral unfoldings of other convex polyhedra,
seeing if they avoid overlap in the plane, i.e., if they unfold to simple (non-self-intersecting) polygons.
First we add some detail to the definition of a spiral.
We insist that a spiral $\s$
satisfies these requirements:
\begin{enumerate}
\squeezelist
\item $p_{i+1}$ is vertically higher or the same height as $p_i$.
Letting $z_i$ be the $z$-coordinate of $p_i$, this condition is $z_{i+1} \ge z_i$.
Therefore, $p_1$ is a bottommost vertex and $p_m$ a topmost vertex of $P$.
\item Each segment $p_i p_{i+1}$ ``advances" counterclockwise (ccw)
around its band.
\end{enumerate}
We defer the somewhat technical definition of what constitutes a
ccw advance to Section~\secref{Bands} below.

\section{Platonic Solids}
\seclab{Platonic}
All of the five Platonic solids have nonoverlapping spiral unfoldings:
the tetrahedron
(Figures~\figref{Tetra3d}
and~\figref{TetraLay}),
\begin{figure}[htbp]
\centering
\begin{minipage}{.48\textwidth}
\centering
\includegraphics[width=0.9\linewidth]{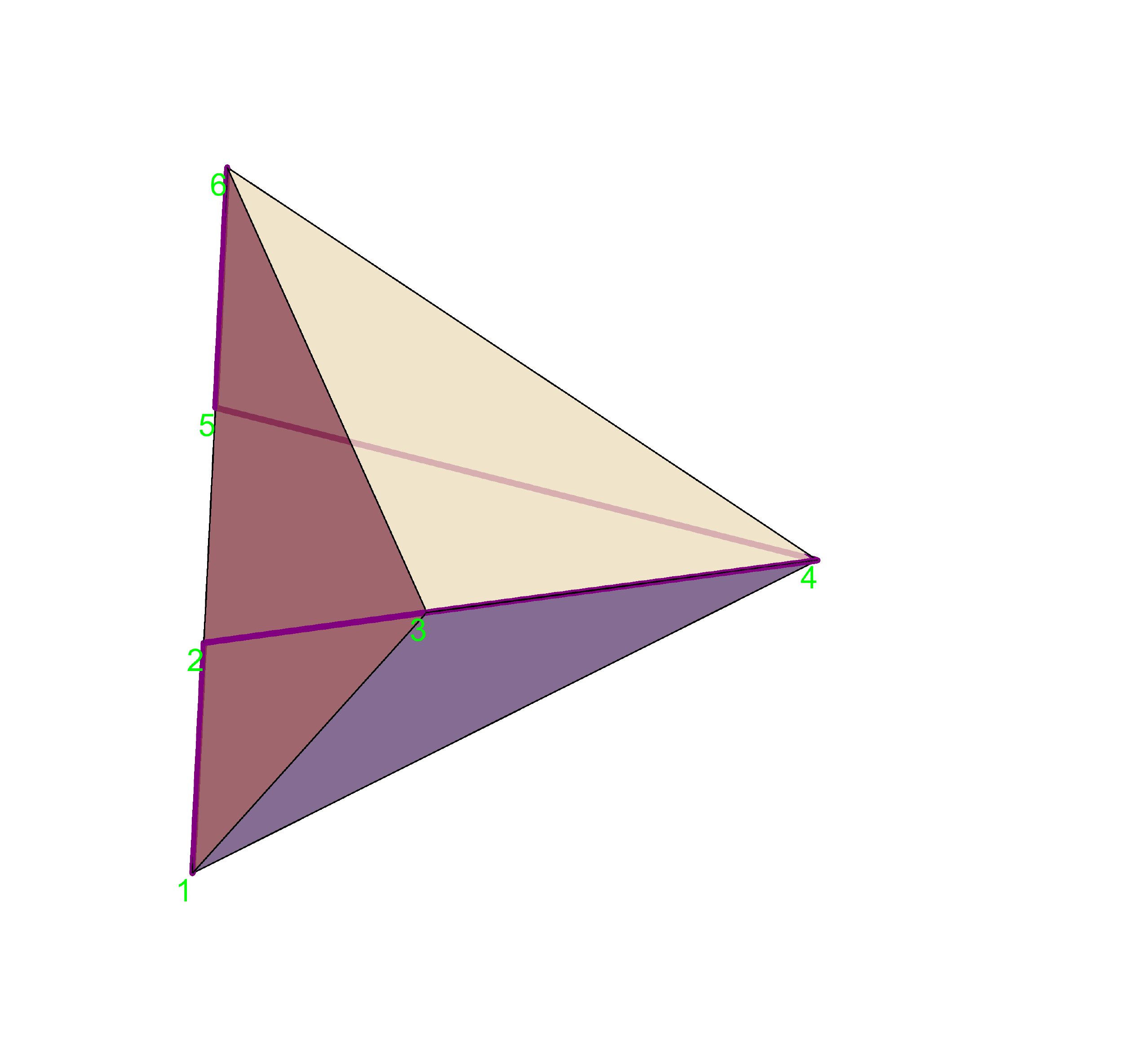}
\caption{Tetrahedron with spiral cut path.
}
\figlab{Tetra3d}
\end{minipage}%
\quad%
\begin{minipage}{.48\textwidth}
\centering
\includegraphics[width=0.8\linewidth]{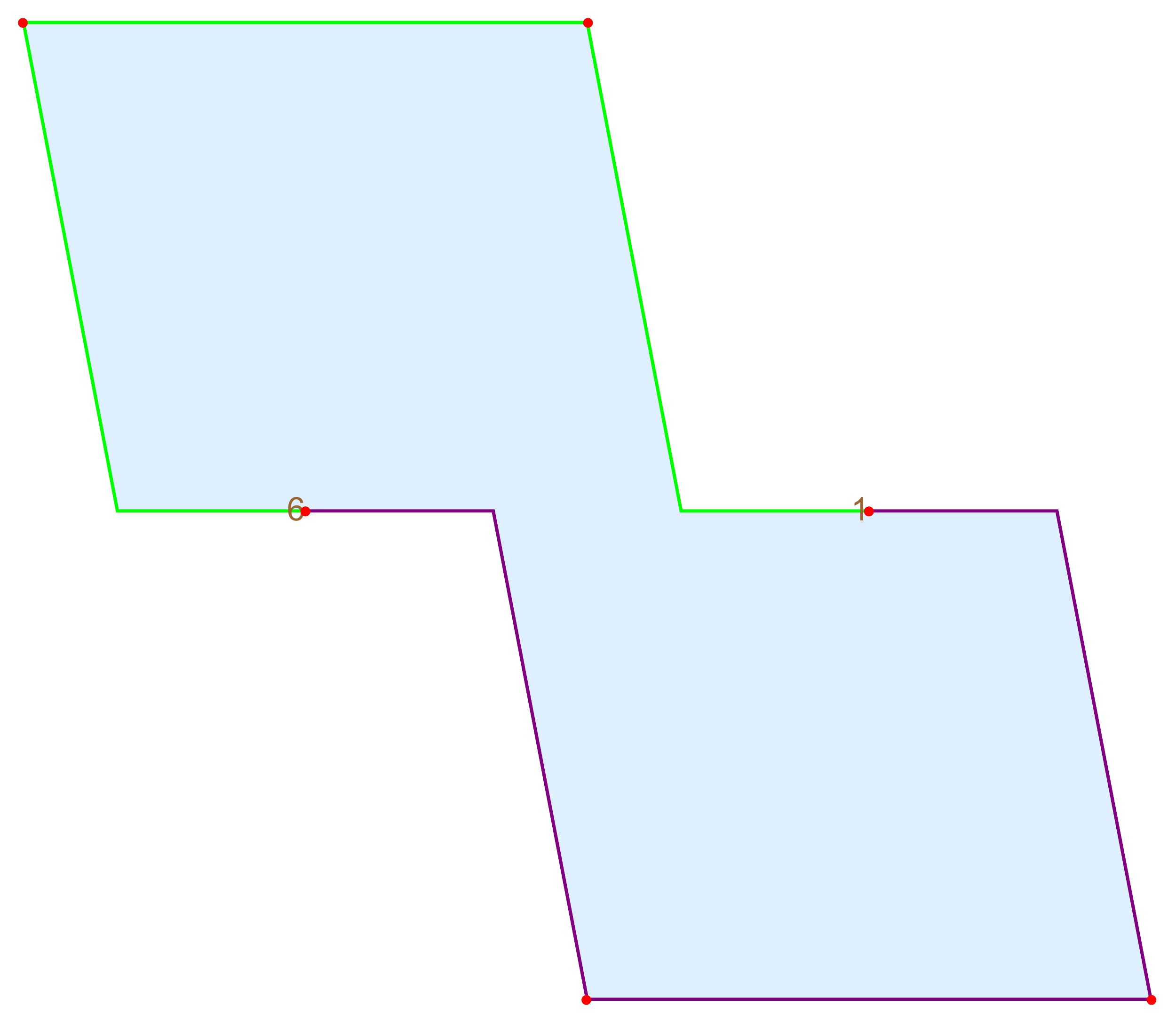}
\caption{Spiral unfolding of tetrahedron.
}
\figlab{TetraLay}
\end{minipage}
\end{figure}
the octahedron (Figures~\figref{Oct3d}
and~\figref{OctLay}),
\begin{figure}[htbp]
\centering
\begin{minipage}{.48\textwidth}
\centering
\includegraphics[width=0.9\linewidth]{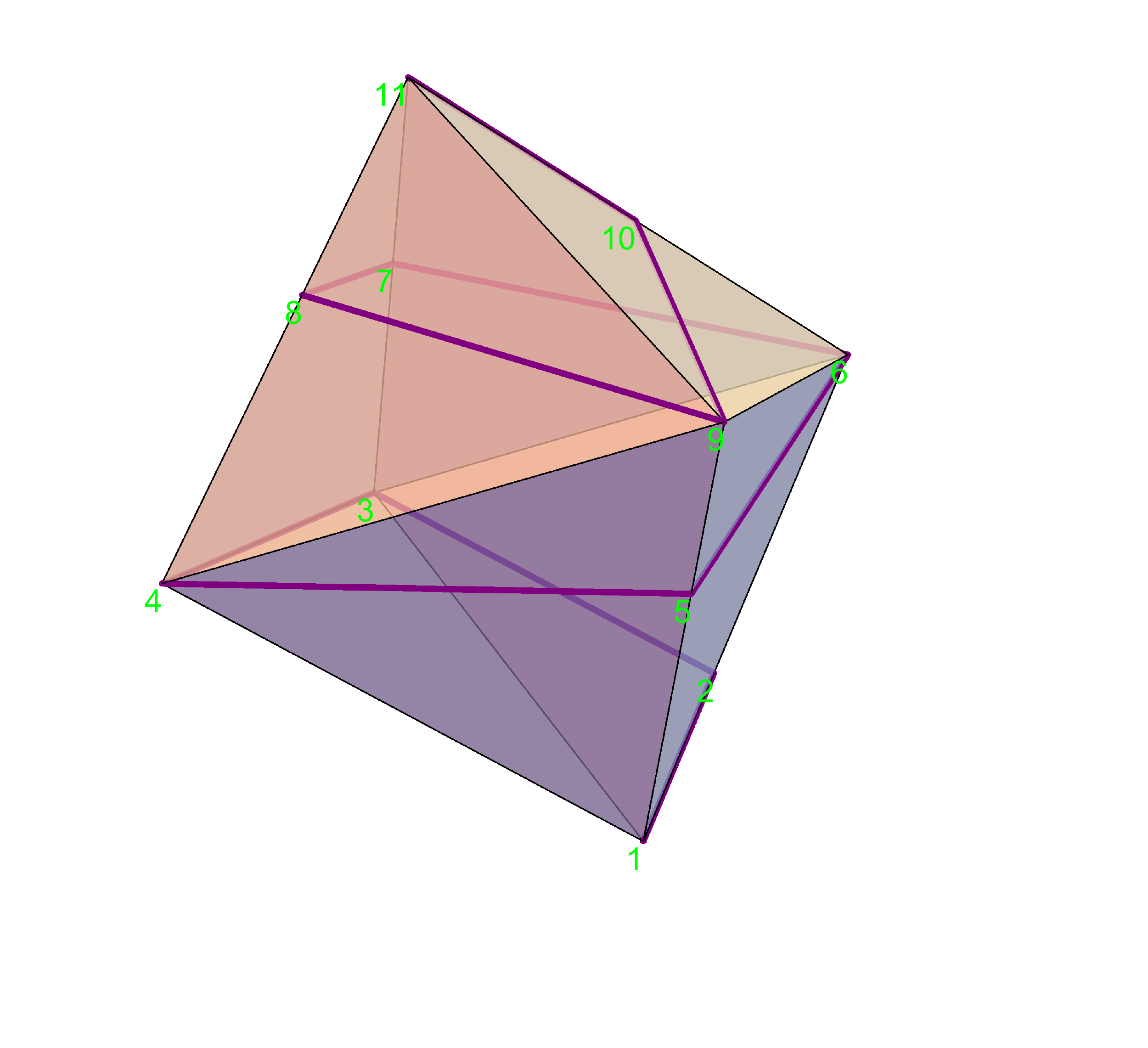}
\caption{Octahedron with spiral cut path.
}
\figlab{Oct3d}
\end{minipage}%
\quad%
\begin{minipage}{.48\textwidth}
\centering
\includegraphics[width=1.0\linewidth]{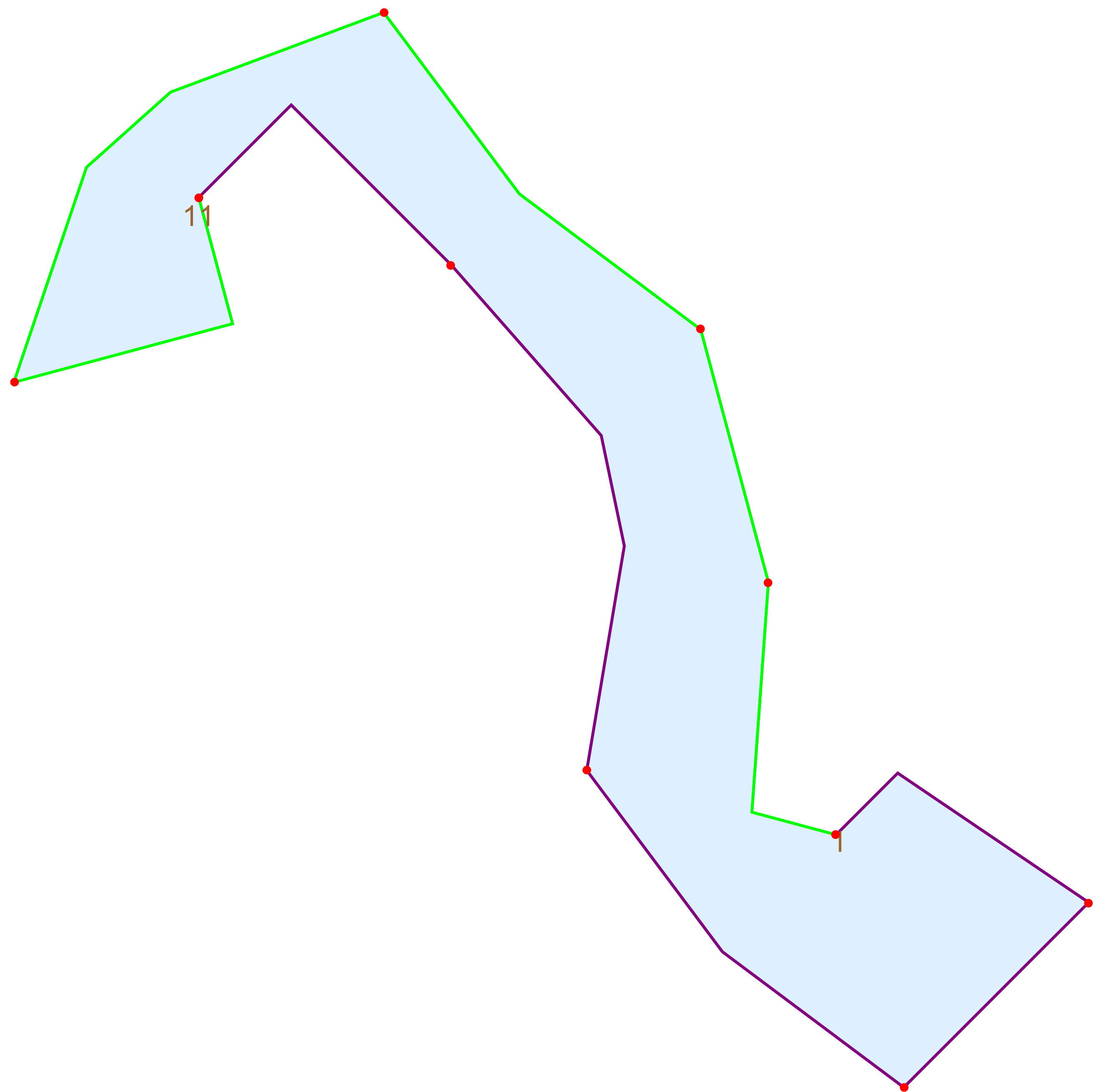}
\caption{Spiral unfolding of octahedron.
}
\figlab{OctLay}
\end{minipage}
\end{figure}
the dodecahedron (Figures~\figref{Dodeca3d}
and~\figref{DodecaLay}),
\begin{figure}[htbp]
\centering
\begin{minipage}{.48\textwidth}
\centering
\includegraphics[width=0.9\linewidth]{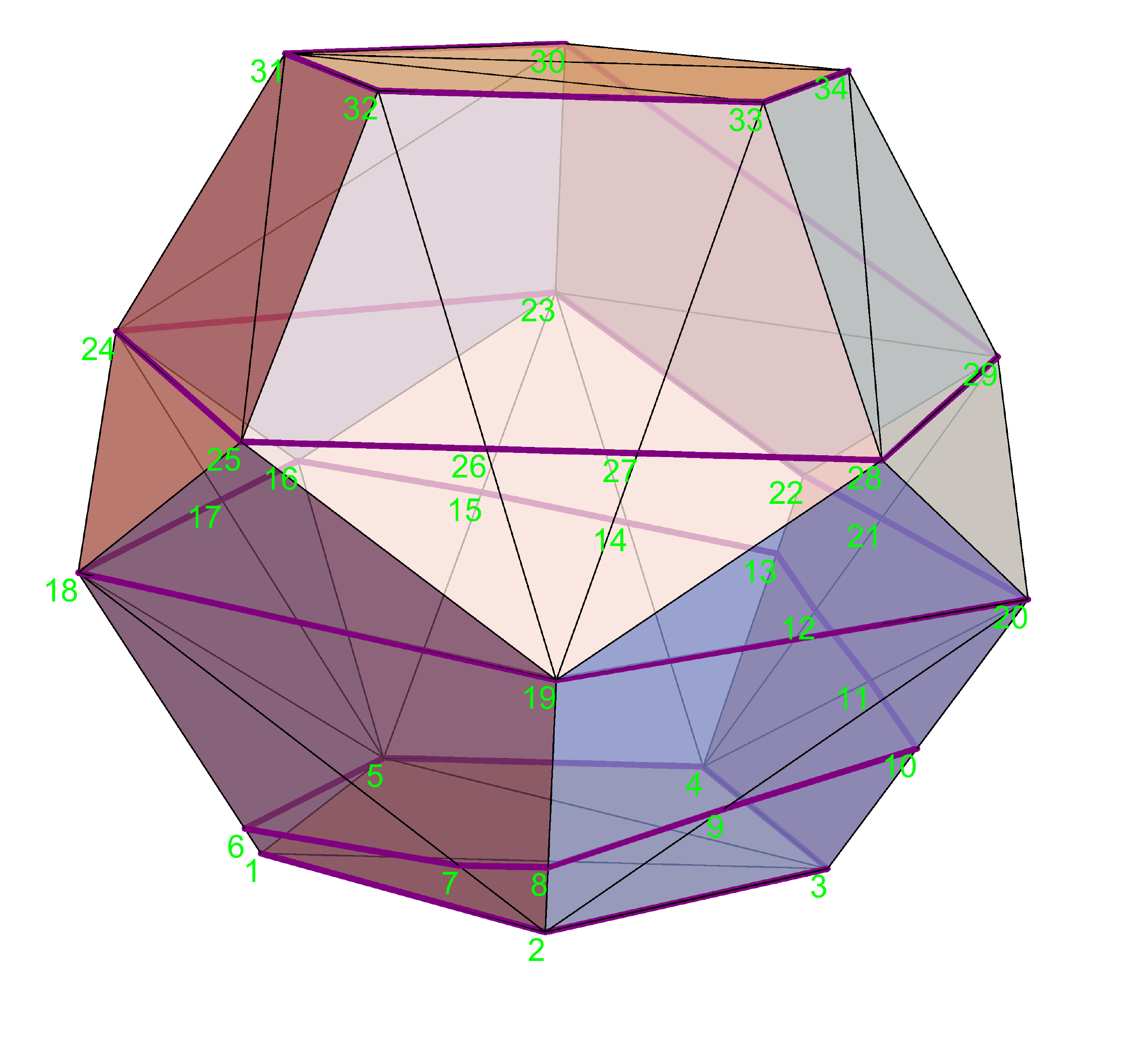}
\caption{Dodecahedron with spiral cut path.
}
\figlab{Dodeca3d}
\end{minipage}%
\quad%
\begin{minipage}{.48\textwidth}
\centering
\includegraphics[width=1.0\linewidth]{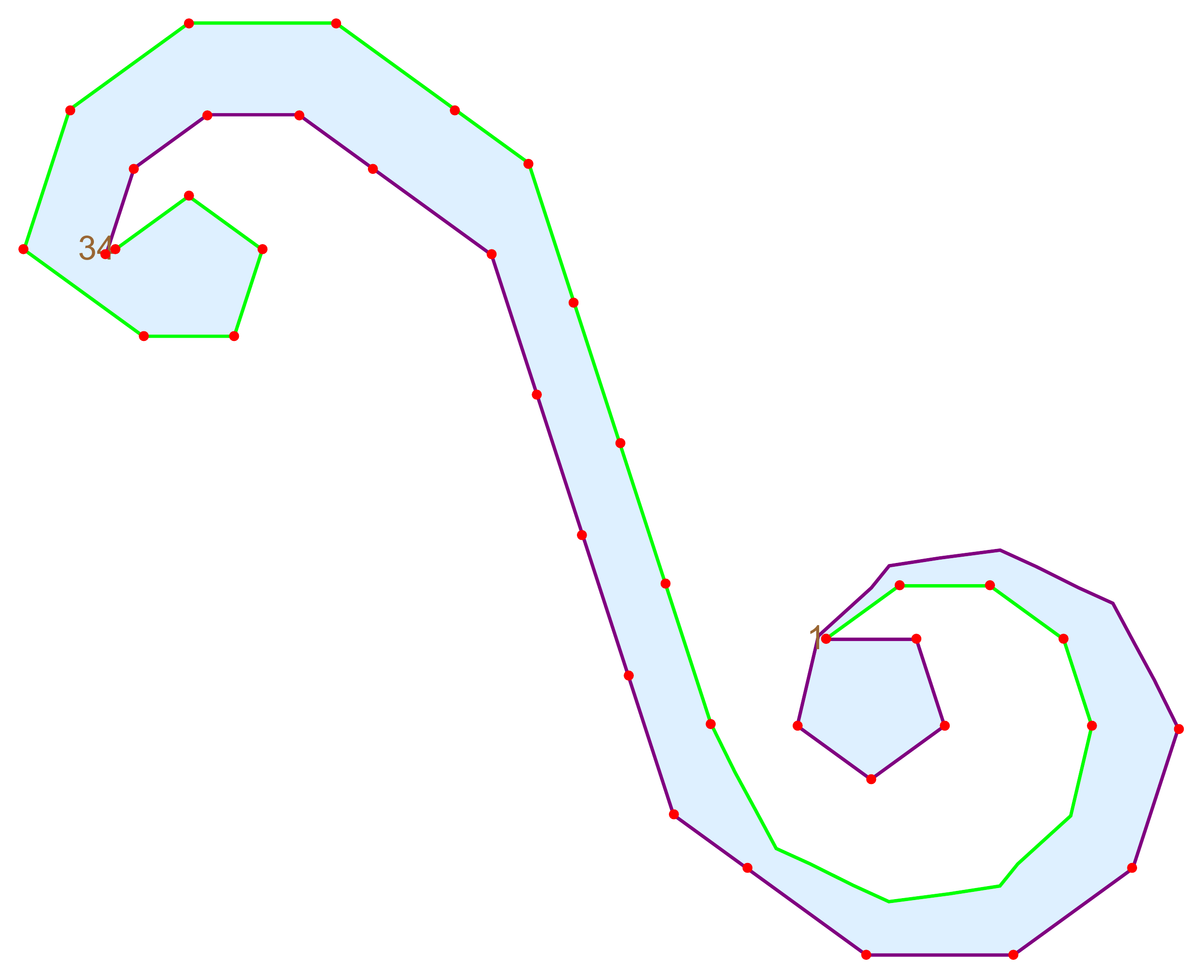}
\caption{Spiral unfolding of dodecahedron.
}
\figlab{DodecaLay}
\end{minipage}
\end{figure}
and the icosahedron (Figure~\figref{Icosa3d}
and~\figref{IcosaLay}).
\begin{figure}[htbp]
\centering
\begin{minipage}{.48\textwidth}
\centering
\includegraphics[width=0.9\linewidth]{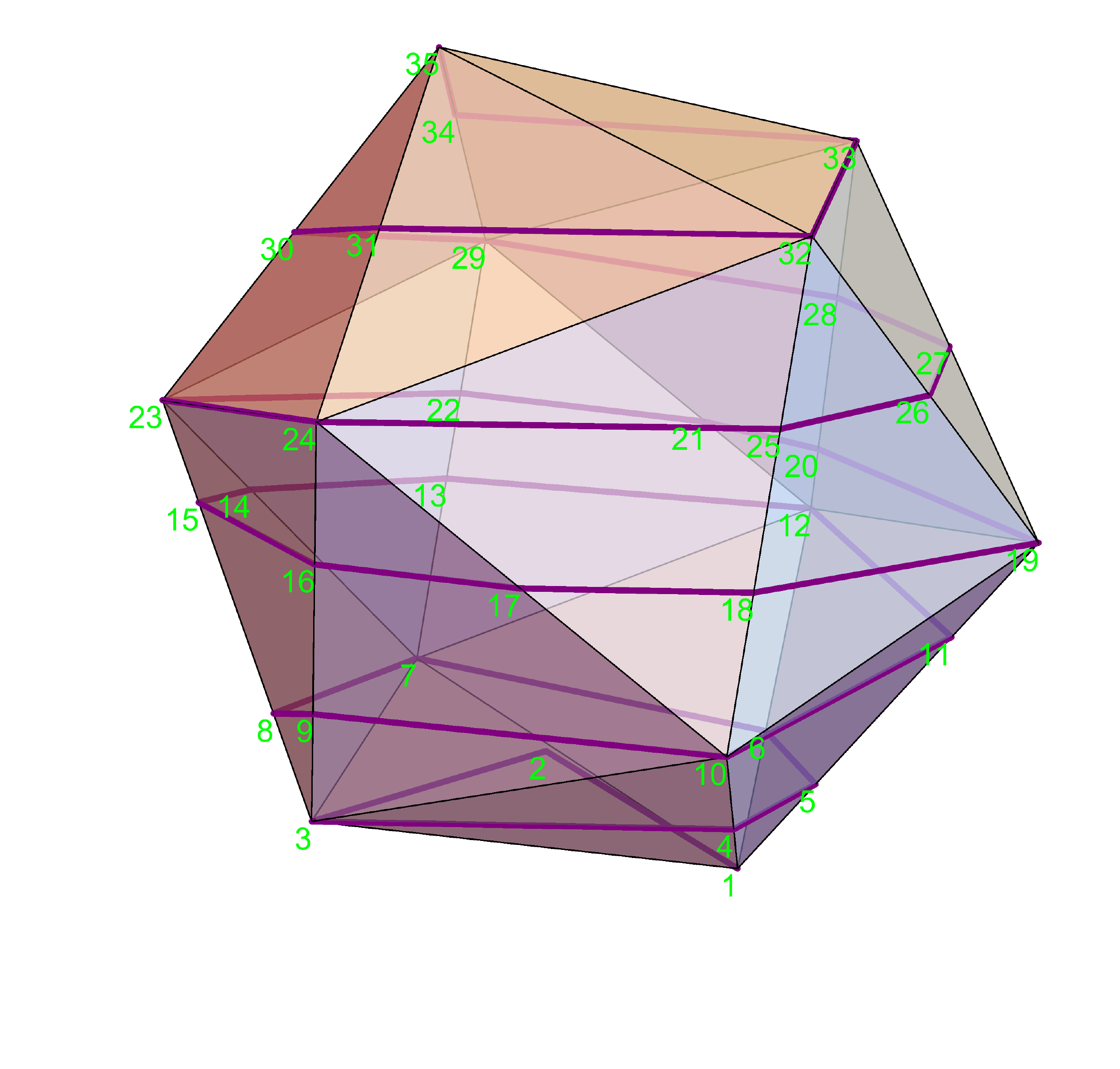}
\caption{Icosahedron with spiral cut path.
}
\figlab{Icosa3d}
\end{minipage}%
\quad%
\begin{minipage}{.48\textwidth}
\centering
\includegraphics[width=1.0\linewidth]{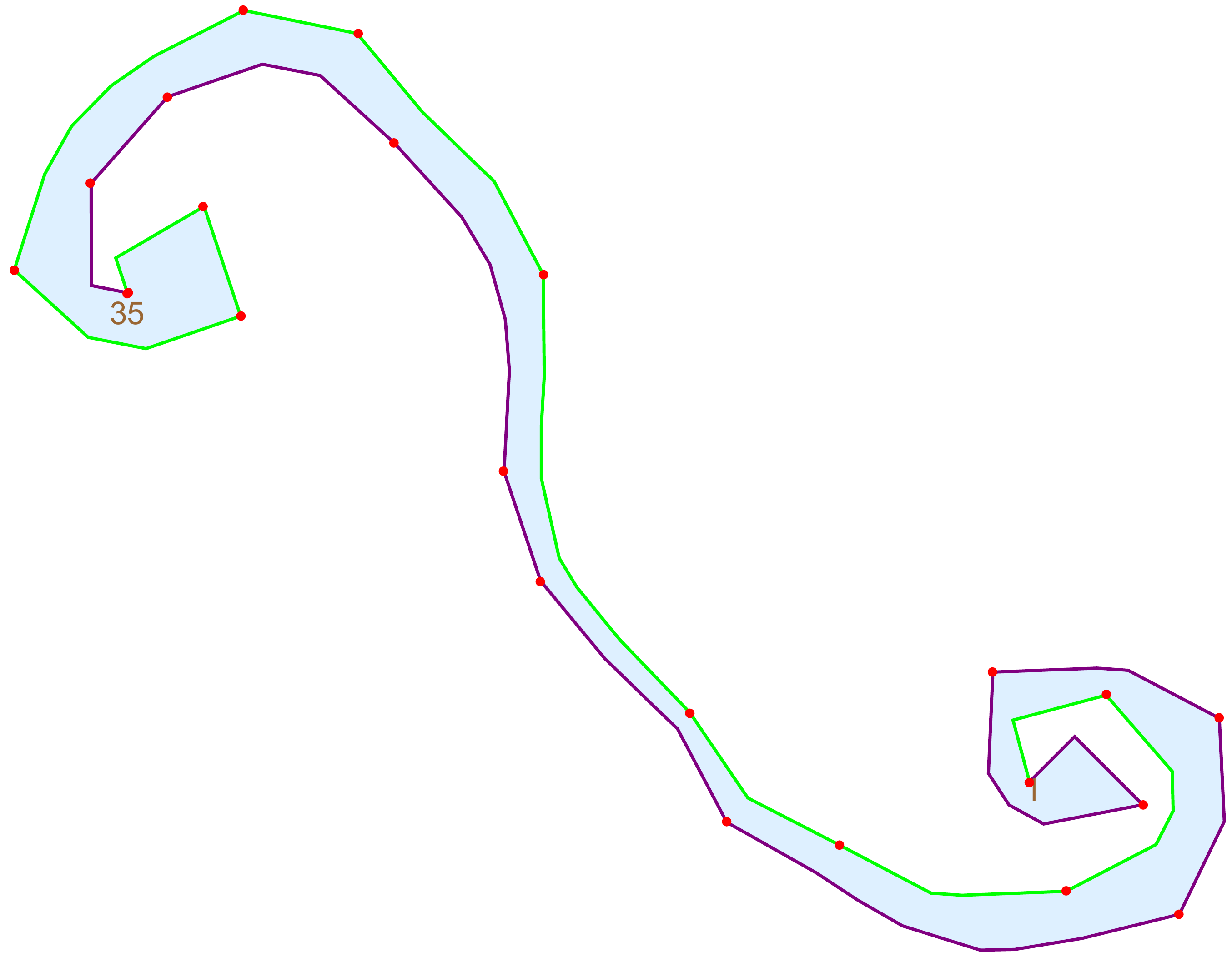}
\caption{Spiral unfolding of icosahedron.
}
\figlab{IcosaLay}
\end{minipage}
\end{figure}

Before proceeding further, a few remarks are in order.
By no means are spirals uniquely defined.
First, the polyhedron may be oriented with two degrees of continuous freedom.
The orientation of the dodecahedron above was carefully selected to avoid overlap:
see Figure~\figref{DodecaOverlap}.
\begin{figure}[htbp]
\centering
\includegraphics[width=0.75\linewidth]{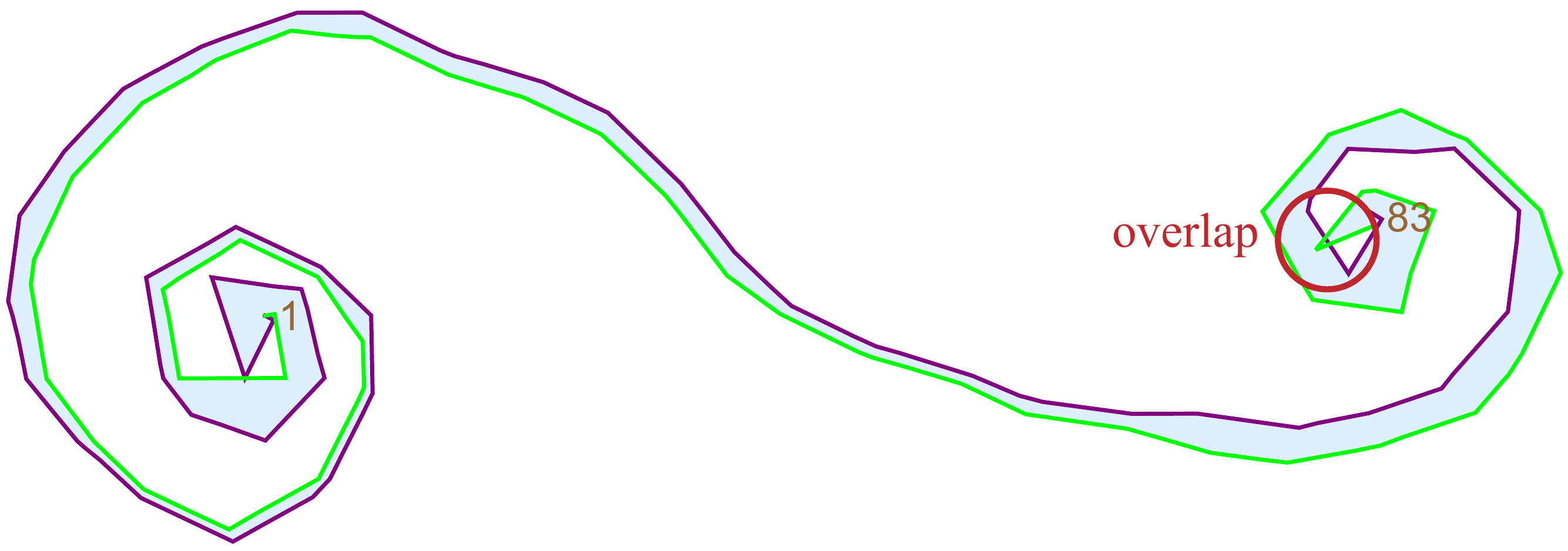}
\caption{Overlapping spiral unfolding of a tilted dodecahedron.}
\figlab{DodecaOverlap}
\end{figure}
Second, for a fixed orientation, even though $\s$ must pass through every vertex
in sorted vertical order, there are still an infinite number of choices for
spirals. For example, one may wind around several times between each vertically
adjacent pair of vertices of $P$;
see Figure~\figref{Octaw2}.
\begin{figure}[htbp]
\centering
\includegraphics[width=0.75\linewidth]{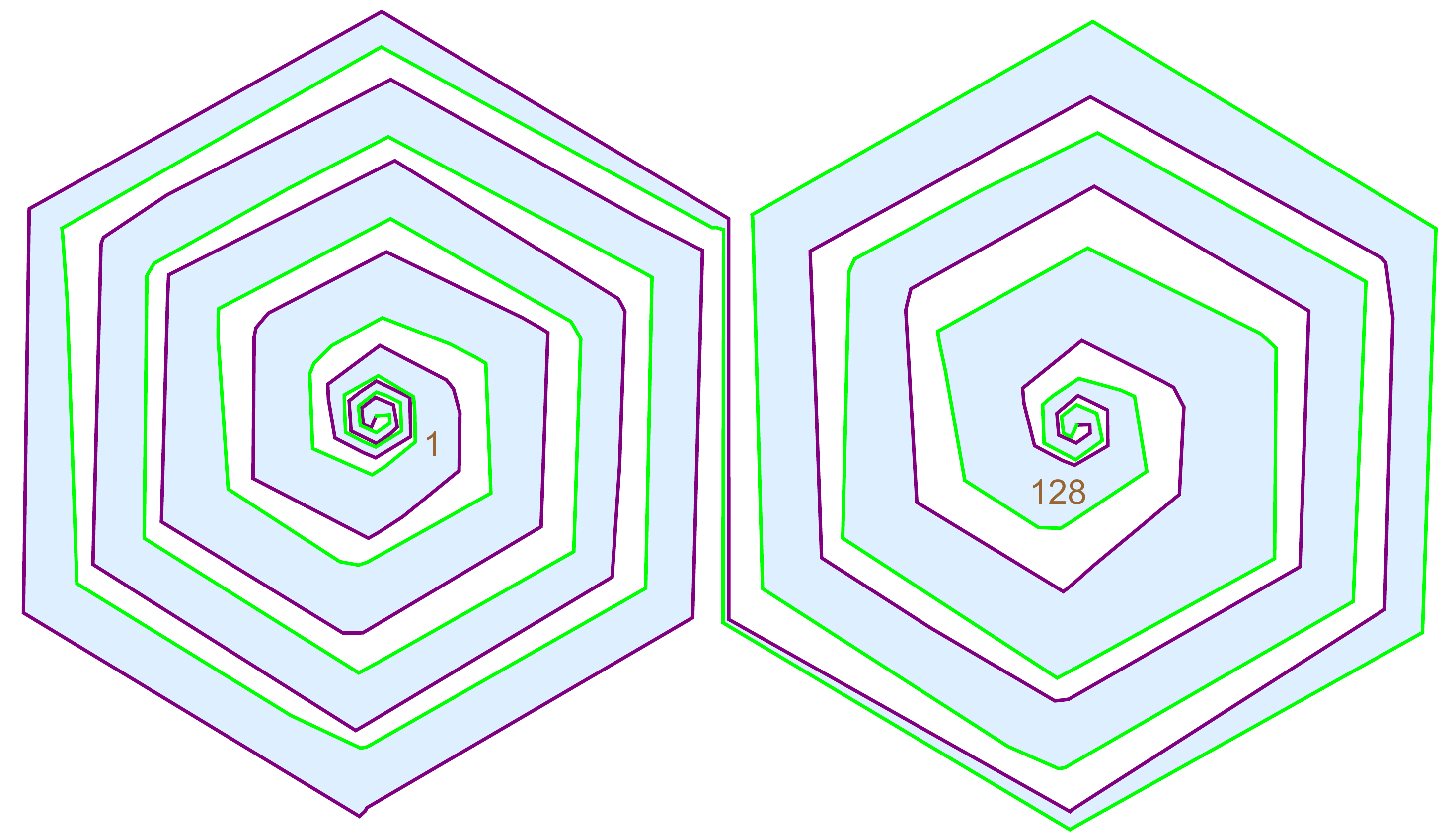}
\caption{Upright octahedron with densely wound spiral cut-path.}
\figlab{Octaw2}
\end{figure}

Further intuition may be gained from an animation
of the previously shown (Figure~\figref{Icosa3d}
and~\figref{IcosaLay}) icosahedron unfolding:
see  Figure~\figref{IcosaF35}.
\begin{figure}[htbp]
\centering
\includegraphics[width=1.0\linewidth]{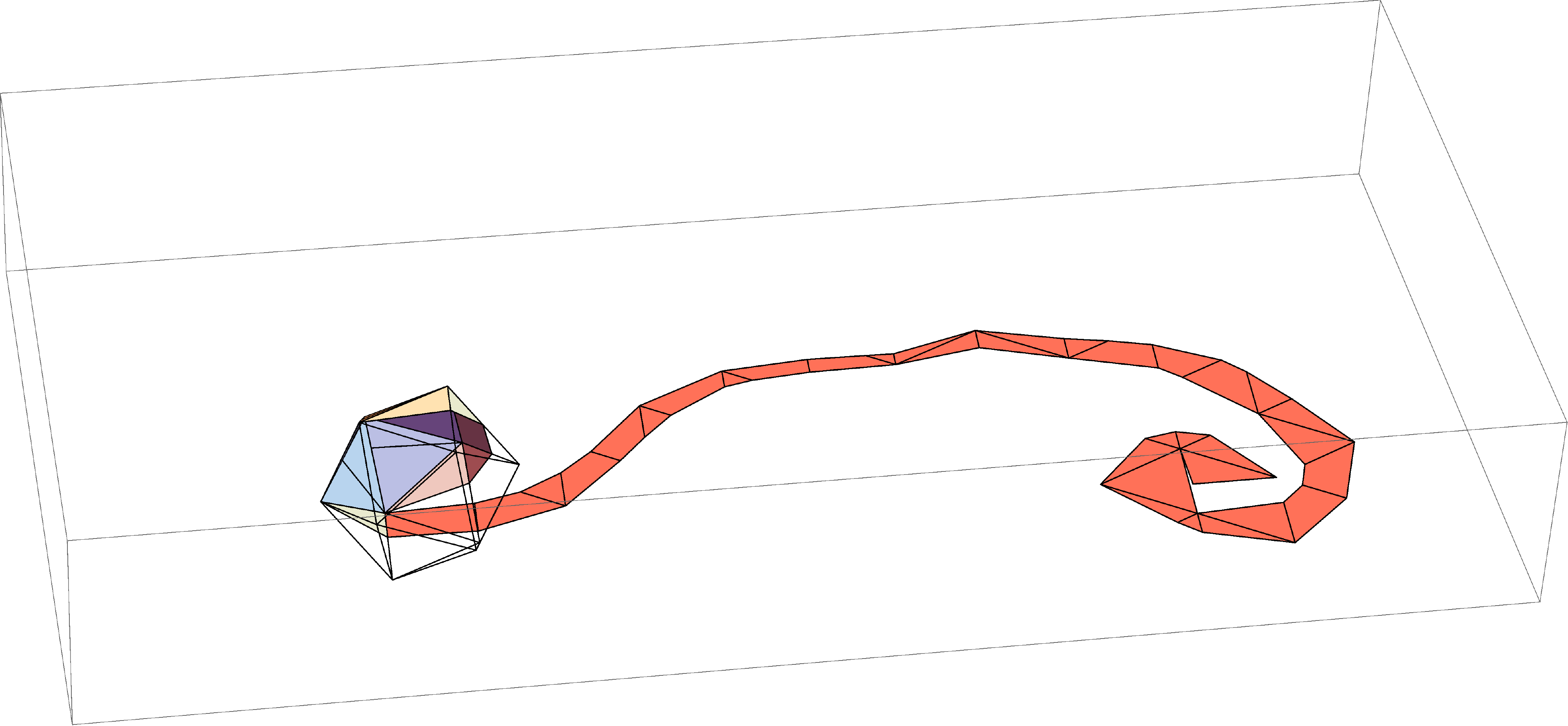}
\caption{Snapshot of animation of icosahedron rolling out its spiral
($35^\text{th}$ out of $50$ frames).
\protect\url{http://cs.smith.edu/~orourke/SpiralUnf/}.
}
\figlab{IcosaF35}
\end{figure}

\subsection{Spiral Definition: Bands}
\seclab{Bands}
We return to the definition of a spiral, to make precise the sense in which the spiral
must always advance ccw around the polyhedron $P$.
Let a \emph{band} be the portion of $P$ between two horizontal
($z=\text{constant}$)
planes, such that there is no vertex of $P$ strictly between the planes
(although vertices may lie on those planes).
See Figure~\figref{Band3D}.
The edges of $P$ crossing the band provide a natural combinatorial sense
of ccw.
If a segment $p_i p_{i+1}$ of $\s$ connects two edges of a band,
then it is ccw if the second edge is ccw of the first edge
(recall our convention that the spiral rises from a bottom to a top vertex).
In general, and in all examples in this paper,
each $p_i$ does in fact lie on an edge (or several, where they meet at
a vertex).
However, this is not a necessary condition: $p_i$ could lie interior to a face. 

If $p_i p_{i+1}$ does not connect two band edges
(e.g., both $p_i$ and $p_{i+1}$ might be 
vertices connected by an edge of $P$), then the following
rule is used to determine whether it represents a ccw advance.
Let $\rho^h_i$ be the total surface angle to the right from $p_i p_{i+1}$ 
down to the
horizontal plane through $p_i$, and similarly let $\lambda^h_i$
be the surface angle to the left. Here the superscript $h$ indicates an
angle to the horizontal plane. For $p_{i+1}$ the analogous angles measure up
to the horizontal plane through $p_{i+1}$.
Then $p_i p_{i+1}$ is ccw iff (a)~$\rho^h_i < \lambda^h_i$,
and (b)~$\rho^h_{i+1} > \lambda^h_{i+1}$;
see Figure~\figref{SpiralCcw}.
This condition requires appropriate ``slant" at each end of $p_i p_{i+1}$,
and ensures vertical symmetry: reversing $z$-coordinates of a path
renders it a spiral iff the original is a spiral.
\begin{figure}[htbp]
\centering
\begin{minipage}{.48\textwidth}
\centering
\includegraphics[width=1.2\linewidth]{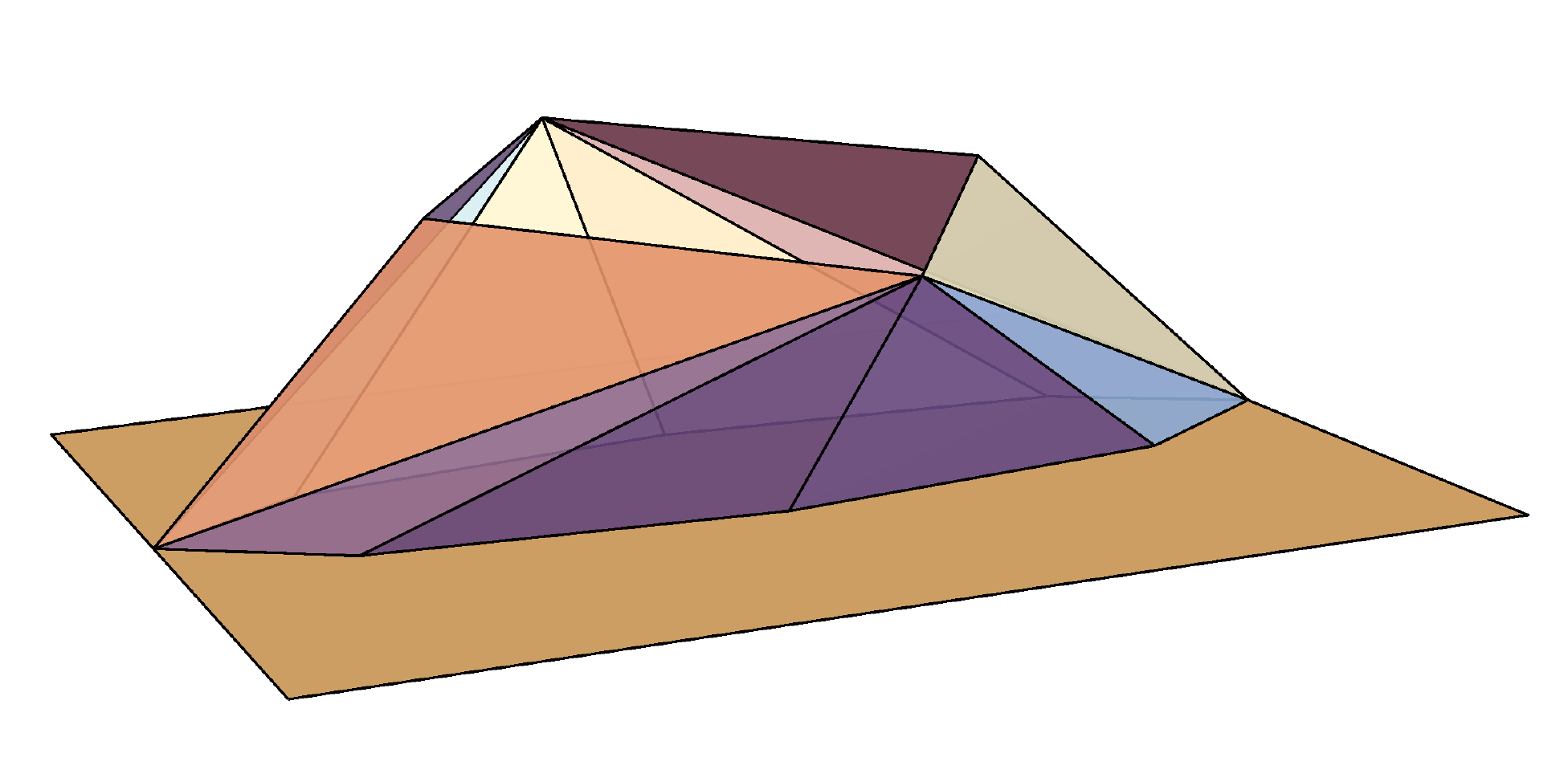}
\caption{A band between two horizontal planes; bottom plane
shown. Both planes pass through vertices in this example.
}
\figlab{Band3D}
\end{minipage}%
\quad%
\begin{minipage}{.48\textwidth}
\centering
\includegraphics[width=0.9\linewidth]{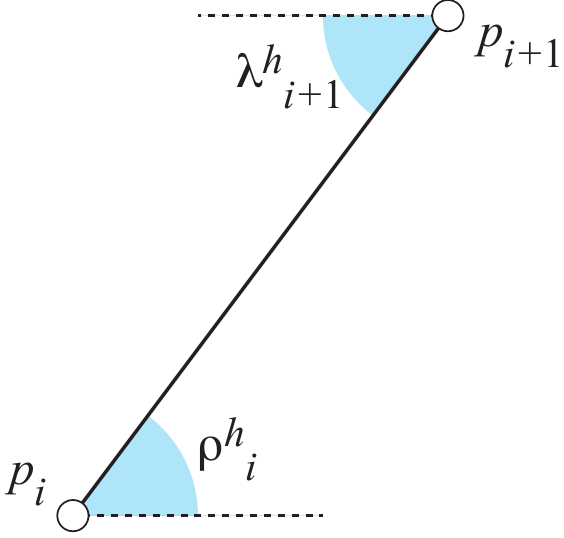}
\caption{Surface angles above and below horizontal planes through
$p_i$ and $p_{i+1}$.
}
\figlab{SpiralCcw}
\end{minipage}
\end{figure}


Before proceeding further, we describe the implementation that produced the unfoldings
displayed in this paper.

\section{Implementation}
\seclab{Implementation}
The spiral cut-paths illustrated were created by a particular implementation,
selecting a specific $\s$ among the infinite number of choices for a given, fixed orientation of $P$.
We first describe the algorithm in the case when no two vertices of $P$ lie at the same height.
At any one time, the portion of $\s$ below and up to a vertex $v_i$ of $P$ has been constructed.
$P$ is sliced with a horizontal plane through $v_i$, and again sliced through the next vertex
in the $z$-direction, $v_{i+1}$.
See Figure~\figref{SlicesBandPath}.
\begin{figure}[htbp]
\centering
\includegraphics[width=0.5\linewidth]{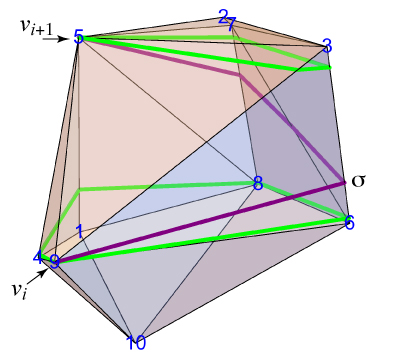}
\caption{The portion of $\s$ between $v_i$ $(9)$ and $v_{i+1}$ $(5)$.
The green polygons outline the intersection 
of $P$ with parallel horizontal planes through $v_i$ and $v_{i+1}$.
}
\figlab{SlicesBandPath}
\end{figure}
These two planes define a vertex-free band $B_i$.
The algorithm takes a parameter $w$ indicating how many complete times $\s$ should wind around $B_i$
before connecting to $v_{i+1}$. In the figure, $w=0$, so the connection is as direct as possible.
In this instance, $v_i v_{i+1}$ is an edge of $P$ and could be followed, but it is not slanting ccw
(Section~\secref{Bands}). Therefore $\s$ spirals across several edges of $P$ before reaching $v_{i+1}$.
Note that, in general, $\s$ cannot follow a geodesic as it winds around the band, as the shape of
a band would only allow that in special circumstances..

The top and bottom vertices are handled specially. Because our goal is to avoid overlap,
and overlap usually occurs at these apexes when acute angles are forced,
the first segment from the bottom vertex $v_1$, to $v_2$,
is selected to make a nonacute turn to connect to $v_2$. This can be seen clearly in Figure~\figref{Cube3d},
where a $90^\circ$ turn is selected,
and in other figures. I proved that such a nonacute turn is always possible.
Of course the turn at $v_2$ to connect to $v_3$ might be highly acute, often
leading to overlap (e.g., in Figure~\figref{DodecaOverlap}). 

Because the algorithm progresses from the bottom vertex to the top without look-ahead,
it is possible that the details of the connection to the top vertex are not identical to the
details at the bottom vertex, even when $P$ is symmetric. This is again evident in Figure~\figref{Cube3d}.

When $P$ is oriented so that more than one vertex lies at a particular $z$-height,
it is necessary, by the definition of a spiral, for $\s$ to cycle around the slice polygon
at that height until all the vertices are included, before angling off to the next vertex vertically.
This is evident for the dodecahedron (Figure~\figref{Dodeca3d}), and for all the Archimedian solid
unfoldings in Section~\secref{Archimedian}.

\section{Archimedian}
\seclab{Archimedian}
All $13$ of the Archimedian solids have nonoverlapping spiral unfoldings.
Six unfoldings are shown in 
Figure~\figref{Archimed1to6},
\begin{figure}[htbp]
\centering
\includegraphics[width=0.9\linewidth]{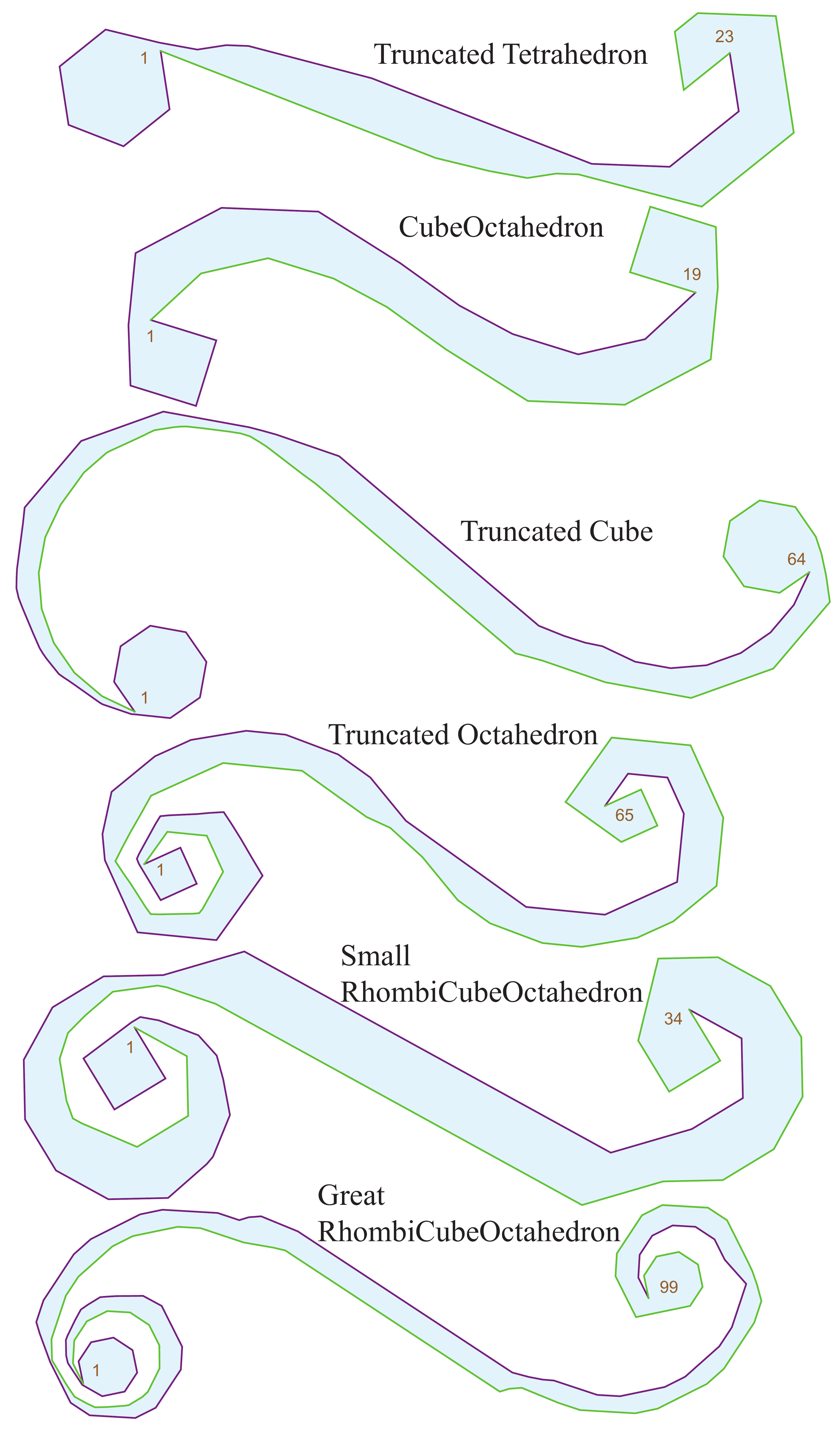}
\caption{Spiral unfoldings of six Archimedian solids.}
\figlab{Archimed1to6}
\end{figure}
six more in Figure~\figref{Archimed7to12},
\begin{figure}[htbp]
\centering
\includegraphics[width=0.9\linewidth]{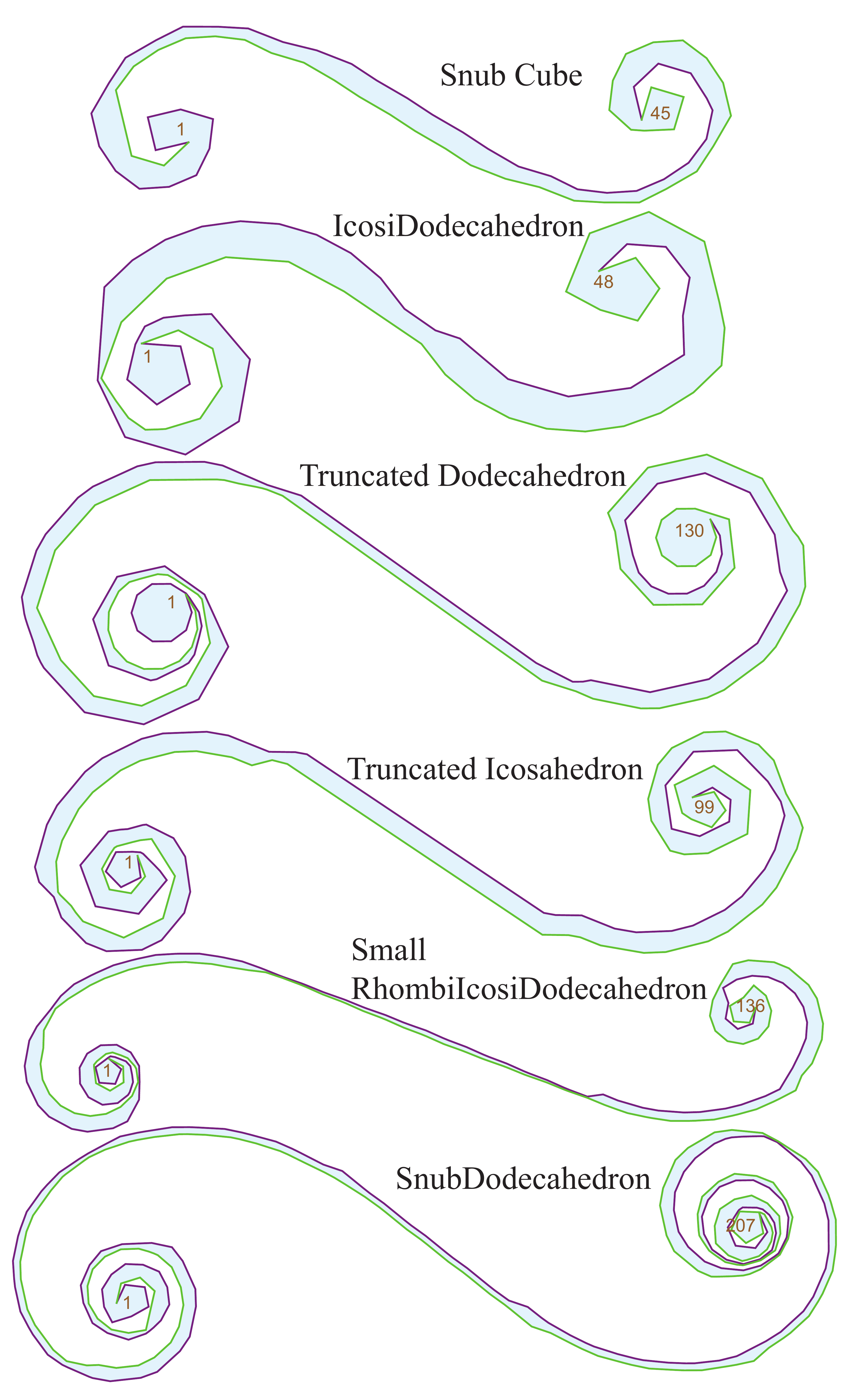}
\caption{Spiral unfoldings of six more Archimedian solids}
\figlab{Archimed7to12}
\end{figure}
and the most complex $13^\text{th}$, the great rhombicosidodecahedron,
is shown in 
Figure~\figref{GRIDLay3D}.
\begin{figure}[htbp]
\centering
\includegraphics[width=0.95\linewidth]{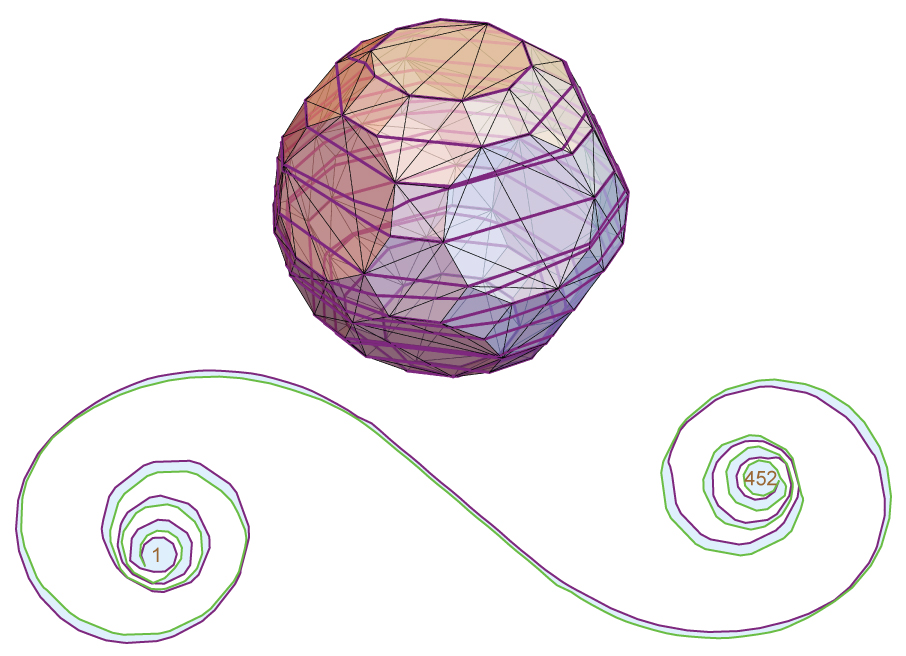}
\caption{Great RhombIcosiDodecahedron.}
\figlab{GRIDLay3D}
\end{figure}
The $\s$ used in this last case has $452$ corners; the polyhedron itself has $60$ vertices.

As shown in~\cite{lddss-zupc-10}, all the Platonic and Archimedian solids have
Hamiltonian edge-unfoldings, and in general they can be chosen to
be ``S-shaped," visually not unlike the spiral unfoldings above.
The exception is the great rhombicosidodecahedron, whose
Hamiltonian edge-unfolding leads to a rather differently shaped planar polygon.

\section{Overlapping spiral unfoldings}
\seclab{Overlapping}
\subsection{Random polyhedra}
\seclab{Random}

Despite the nonoverlapping spiral unfoldings of the Platonic and Archimedian solids,
avoiding overlap is actually rare.
Figure~\figref{PercentOverlap} shows data from polyhedra constructed as
the convex hull of random points uniformly distributed on a sphere.
By the time $P$ has $25$ vertices, essentially no random polyhedron, in a random orientation,
lead to a nonoverlapping unfolding using the spiral cut-path generated by the algorithm
discussed in the previous section.
\begin{figure}[htbp]
\centering
\includegraphics[width=0.75\linewidth]{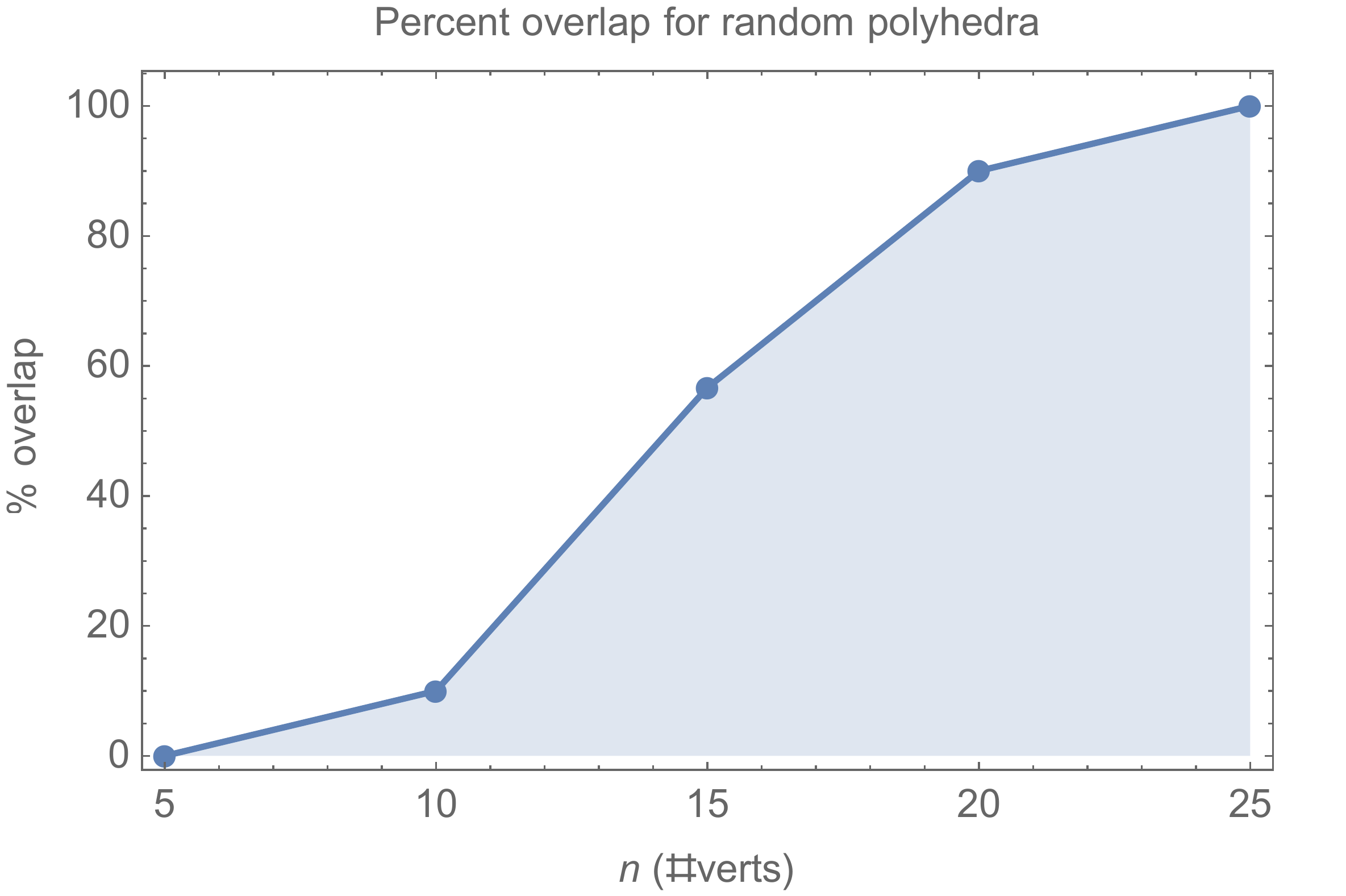}
\caption{The percentage of overlapping spiral unfoldings for random polyhedron with $n$ vertices.
Each point plotted is the mean of $50$ random trials.}
\figlab{PercentOverlap}
\end{figure}
The intuition for why overlap is common can be seen in the nonoverlapping spiral unfolding
of the $60$-vertex great rhombicosidodecahedron in Figure~\figref{GRIDLay3D}:
with many vertices, the spiral turns many times near the apex, and those must
unfold to a precisely nested, tightly wound spiral in the plane to avoid overlap.

Of course, this random data just suggests that overlap is common, not that it cannot be avoided.

\subsection{Polyhedra without nonoverlapping spiral unfoldings}
\seclab{cex}
Despite the rarity of nonoverlapping spiral unfoldings,
it is not straightforward to identify a particular polyhedron $P$ that
has no nonoverlapping spiral unfolding, for two reasons:
(1)~For a given orientation of $P$, there are an infinite number of spirals compatible with that orientation.
(2)~All orientations must be blocked.
Here I propose a $P$ which almost certainly has no spiral unfolding,
but my argument for this falls short of a formal proof.

Define a \emph{hemiball} $H$ as the convex hull of a circle and a semicircle
of equal radii,
as illustrated in 
Figure~\figref{HemiBallRims}.
Here the full (red) circle $C$
lies in the $xy$-plane, and the (green) semicircle $C^+$ lies in the $xz$-plane.
Let $n$ be the number of points equally spaced around $C$, with $n/2$ around $C^+$.
Figure~\figref{HemiBall3D} shows the convex hull $H_n$.
\begin{figure}[htbp]
\centering
\begin{minipage}{.48\textwidth}
\centering
\includegraphics[width=0.9\linewidth]{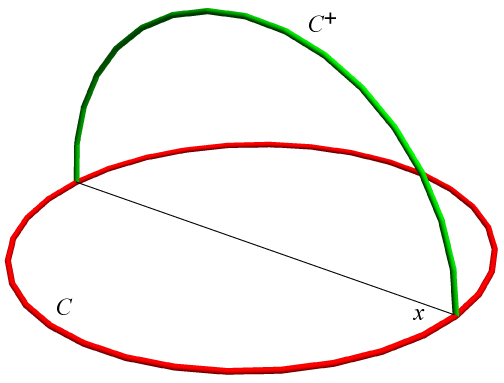}
\caption{HemiBall rims.
}
\figlab{HemiBallRims}
\end{minipage}%
\quad%
\begin{minipage}{.48\textwidth}
\centering
\includegraphics[width=0.9\linewidth]{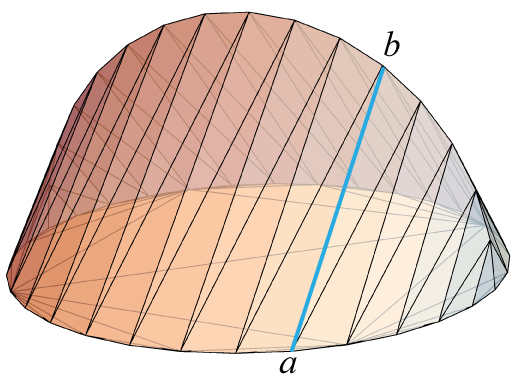}
\caption{HemiBall $H_n$, $n=32$.
}
\figlab{HemiBall3D}
\end{minipage}
\end{figure}
The lateral hull edges $ab$ connect $a(\q)$ on $C$ and  $b(\q)$ on $C^+$, where
\begin{eqnarray}
a(\q) & = & (\cos \q, \sin \q, 0)\;, \\
b(\q) & = & (\cos \q, 0, | \sin \q |)\;.
\end{eqnarray}

\begin{conjecture}
For sufficiently large $n$, the hemiball $H_n$ has no nonoverlapping spiral unfoldings.
\end{conjecture}
I now present evidence for this conjecture, for $H=H_{16}$.
There are two special orientations of $H$.
The first is when the flat base is horizontal; see Figure~\figref{HB0013D}.
Then spiral unfoldings overlap near the top of $H$, as in Figure~\figref{HB001Lay}.
This illustrates the logic of $H$: planes slicing through the rims cut highly
non-circular bands, which tend to lead to acute angles and overlapping unfoldings.
\begin{figure}[htbp]
\centering
\begin{minipage}{.48\textwidth}
\centering
\includegraphics[width=0.9\linewidth]{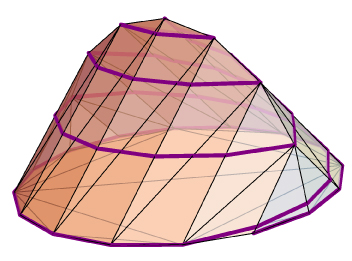}
\caption{$H$ oriented with $(0,0,1)$ vertical.
}
\figlab{HB0013D}
\end{minipage}%
\quad%
\begin{minipage}{.48\textwidth}
\centering
\includegraphics[width=0.9\linewidth]{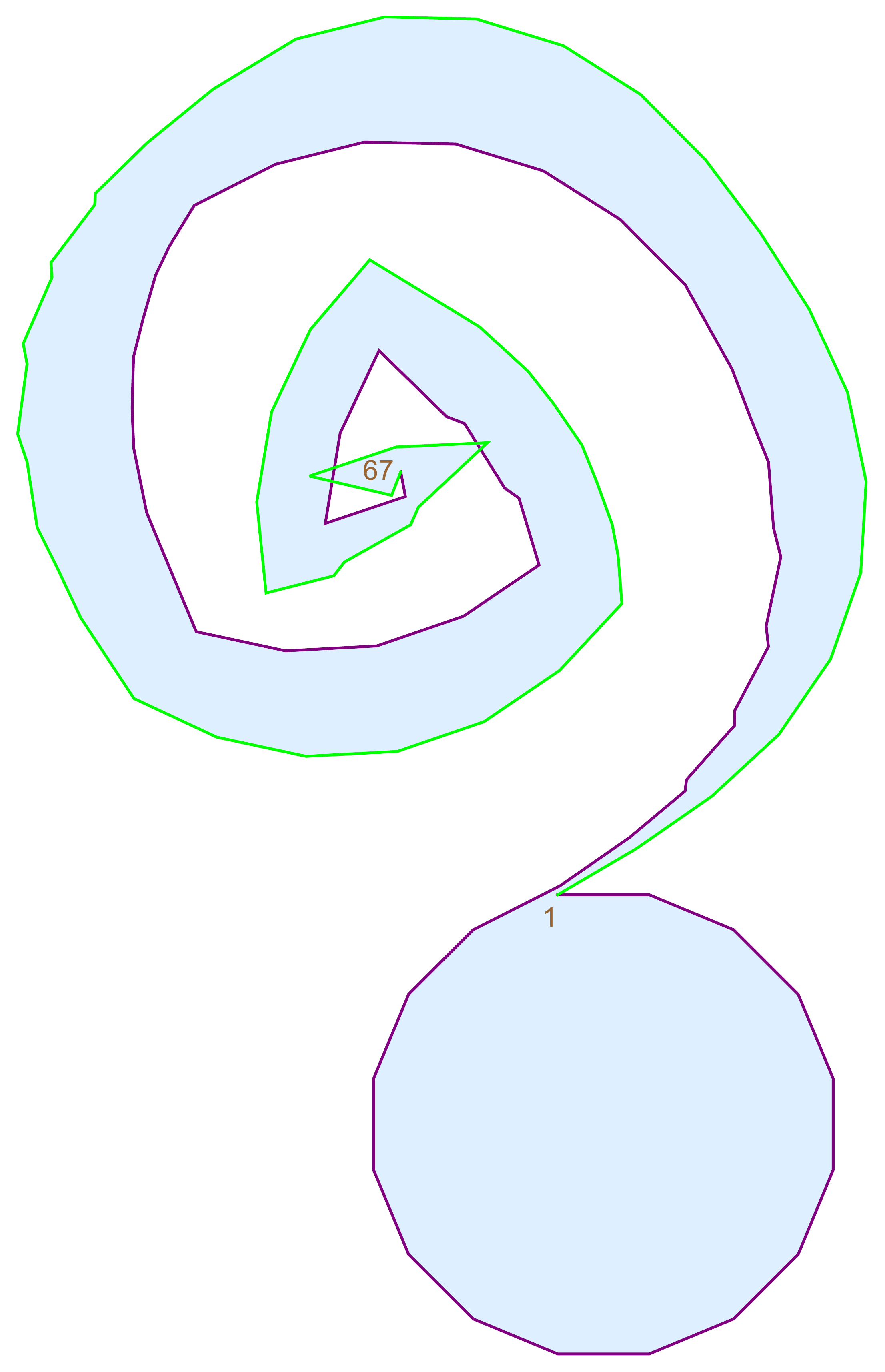}
\caption{Overlapping unfolding.
}
\figlab{HB001Lay}
\end{minipage}
\end{figure}

A second special orientation, tilting $H$ $90^\circ$, is shown in
Figure~\figref{HB1003D}, which leads to overlap at both ends:
Figure~\figref{HB100Lay}.
\begin{figure}[htbp]
\centering
\begin{minipage}{.48\textwidth}
\centering
\includegraphics[width=0.75\linewidth]{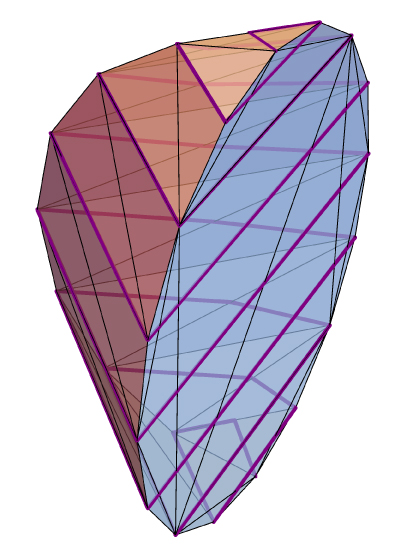}
\caption{$H$ oriented so $(1,0,0)$ is vertical.
}
\figlab{HB1003D}
\end{minipage}%
\quad%
\begin{minipage}{.48\textwidth}
\centering
\includegraphics[width=0.9\linewidth]{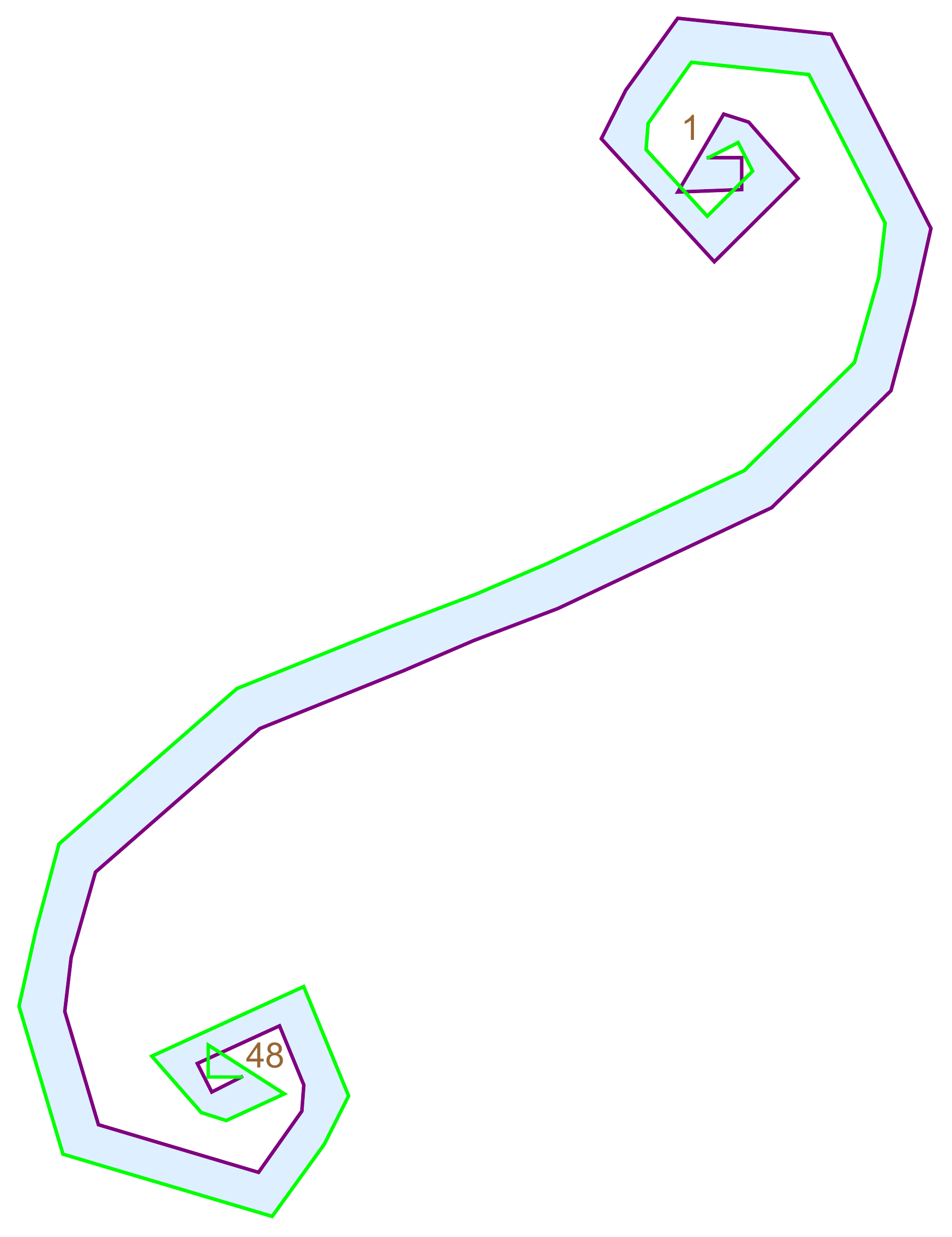}
\caption{Overlapping unfolding.
}
\figlab{HB100Lay}
\end{minipage}
\end{figure}

For other orientations of $H$, either the top or the bottom (or both) plane slices
cut $H$ in eccentric bands, which result in overlaps.
A typical example is shown in  Figure~\figref{HemiBallRandomLay}.
\begin{figure}[htbp]
\centering
\includegraphics[width=0.75\linewidth]{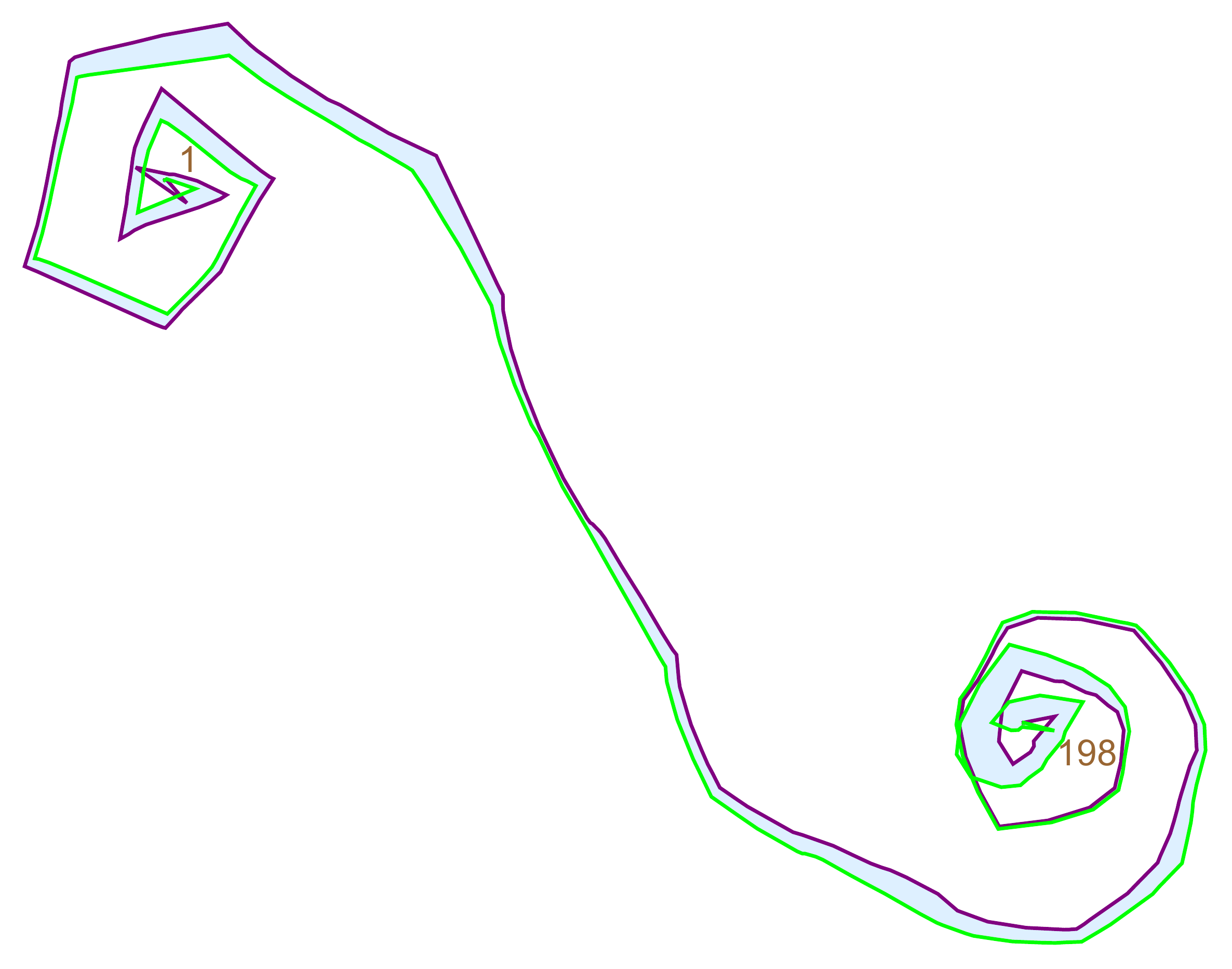}
\caption{Random spiral unfolding of $H$.}
\figlab{HemiBallRandomLay}
\end{figure}
I have verified overlap occurs in hundreds of random orientations of $H$,
which of course only implies that overlap is common, not necessary.
I suggest a proof of necessity might be formulated around the following approach.

Figure~\figref{NormalVecs} illustrates the normal vectors to $H$ in its standard position.
\begin{figure}[htbp]
\centering
\includegraphics[width=0.5\linewidth]{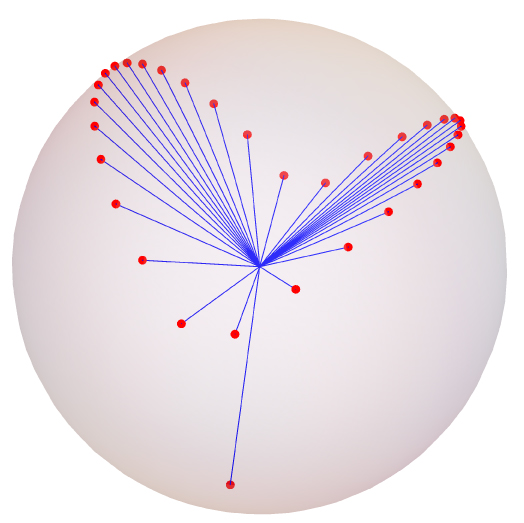}
\caption{Normal vectors to $H$ faces (Figure~\protect\figref{HemiBall3D}).
The single vector in the southern hemisphere
derives from the horizontal base.}
\figlab{NormalVecs}
\end{figure}
Now consider orienting $H$ so that $v$ is normal to the horizontal
slicing planes. We need only
consider $v$ in the northern hemisphere of the figure.
If $v$ lies within the $90^\circ$ wedge containing the northpole,
then these nearly horizontal slicing planes will cut the upper rim $C^+$
in a manner similar to that seen in Figure~\figref{HB0013D}.
In particular, the top vertex will be unique.
If $v$ lies outside of that wedge, then the nearly vertical 
slicing planes will cut the rim $C$, and the bottom vertex will be unique.
In either case, a rim is sliced at one end or the other, and such slices will
be elongated by design, leading to overlapping unfoldings.

I believe this argument could be formalized.
Incidentally, it is easy to find a nonoverlapping unzipping of the hemiball:
see Figure~\figref{HemiBallUnzip}.
\begin{figure}[htbp]
\centering
\includegraphics[width=0.75\linewidth]{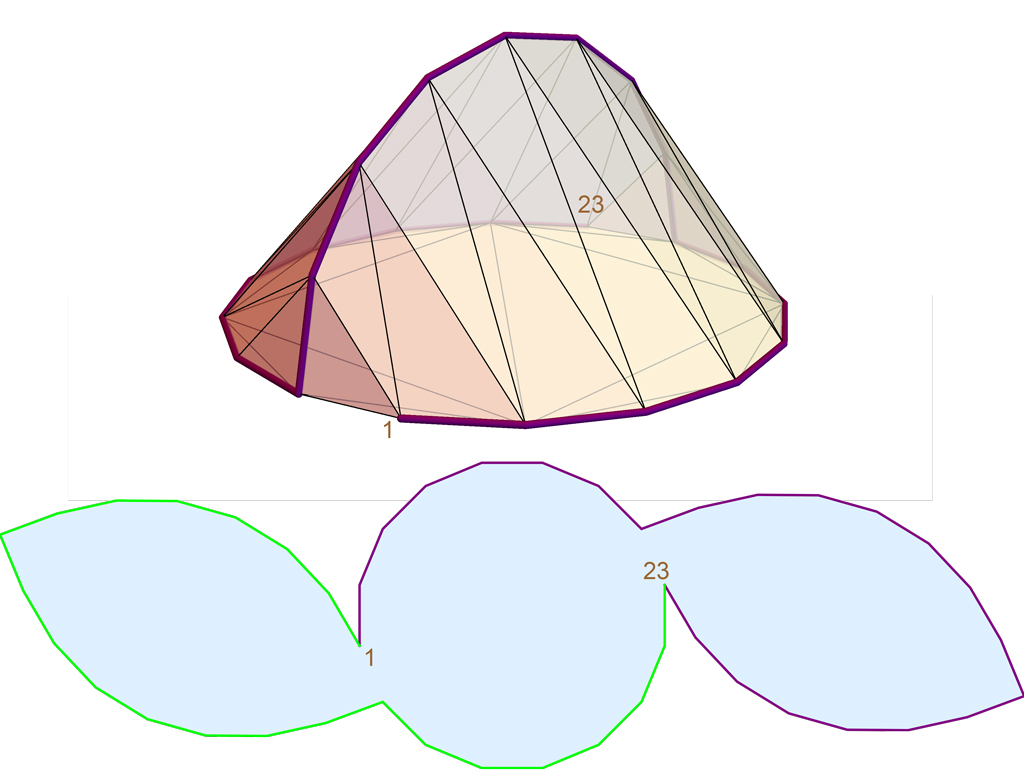}
\caption{A zipper path that unfolds the hemiball $H$.}
\figlab{HemiBallUnzip}
\end{figure}

Another candidate for a polyhedron with no nonoverlapping spiral unfoldings
is a distorted dodecahedron, distorted so that no two pentagonal faces
are parallel.
A flat doubly covered regular polygon with large $n$ has just one orientation
whose spirals avoid overlap: that with the disk horizontal, in which case it
unfolds to two connected regular polygons.

\section{Polyhedra of Revolution}
\seclab{PolyRev}
In this section we study a narrow class of convex polyhedra whose spiral unfoldings
are easily understood.
I believe this sheds light on the general case.
The class is \emph{polyhedra of revolution},
formed as follows.
Let $C$ be a convex curve in the $yz$-halfplane with $x \ge 0$.
Spin this curve around the $z$-axis, forming $n_\text{spin}$ discrete copies.
Then take the convex hull to form $P$.

Two examples for the same $C$ are shown Figure~\figref{PolRevr4n6s4203D}.
\begin{figure}[htbp]
\centering
\includegraphics[width=0.75\linewidth]{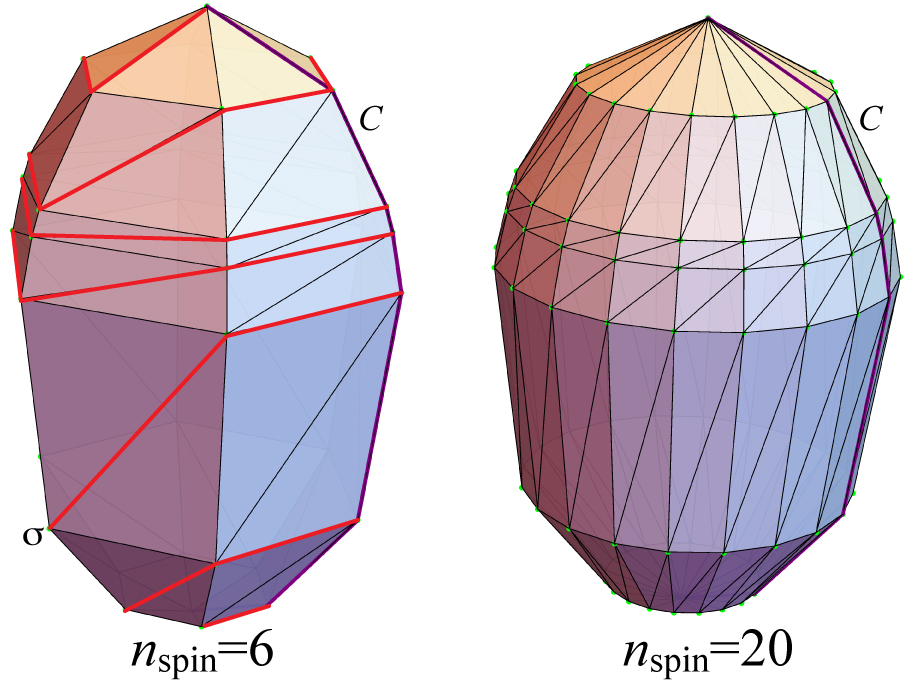}
\caption{Polyhedra of revolution, with 
$C$ shown,
and $n_\text{spin}=4$ and $20$. The spiral cut-path $\s$ is shown on the left polyhedron.}
\figlab{PolRevr4n6s4203D}
\end{figure}
Note the faces of such a $P$ are trapezoids, but in the triangulated figure, each
trapezoid is cut by a diagonal.
Let $v$ be a vertex of $C$.
The natural spiral cut-path $\s$ we explore circles around the 
horizontal regular polygon
of $n_\text{spin}$ sides formed by each $v$, following a trapezoid diagonal
to the next ring upward.\footnote{%
If $\s$ follows a vertical trapezoid edge rather than the coplanar diagonal,
$\s$ would constitute a Hamiltonian edge-unfolding.}

Typical spiral unfoldings are shown in 
Figure~\figref{PolRevr4n6ThreeLay}.
Here we see overlap for $n_\text{spin} \le 4$, just barely nonoverlap for $n_\text{spin}=6$,
and nonoverlap for all larger $n_\text{spin}$.
\begin{figure}[htbp]
\centering
\includegraphics[width=0.75\linewidth]{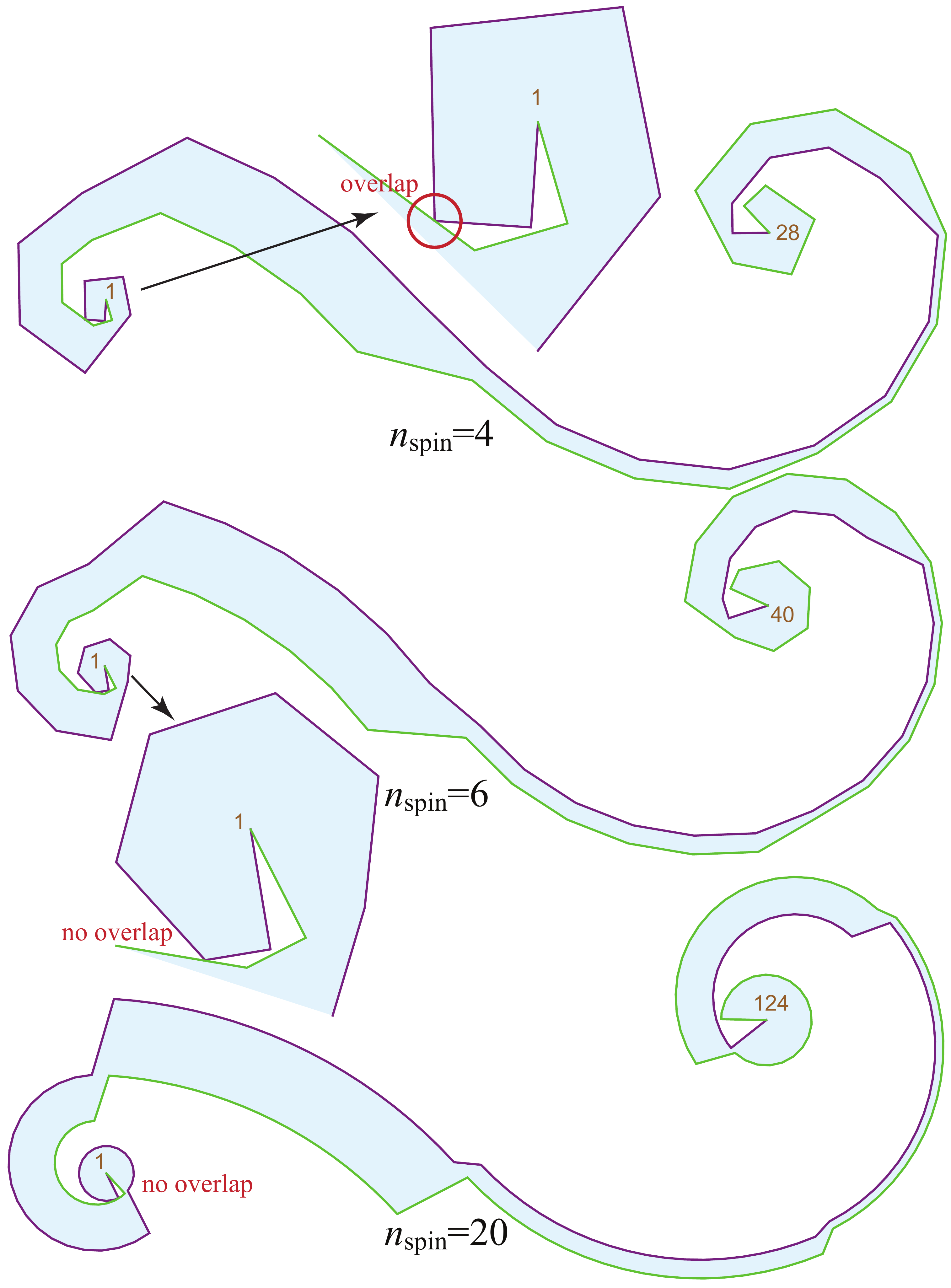}
\caption{Spiral unfoldings of the shape in Figure~\protect\figref{PolRevr4n6s4203D},
for $n_\text{spin}=4,6,20$.}
\figlab{PolRevr4n6ThreeLay}
\end{figure}
The claim of this section is that this is the general situation:

\begin{theorem}
For any convex curve $C$, there is some\footnote{
$n_\text{spin} =2$ produces a flat, doubly covered convex polygon, which
is usually considered a convex polyhedron in the context of unfolding.} 
$n_\text{spin} \ge 2$
such that
the spiral unfolding of the polyhedron of revolution determined
by $C$ and all $\ge n_\text{spin}$ does not overlap.
\thmlab{PolyRev}
\end{theorem}

We now argue for this proposition, somewhat informally.
First, it helps to imagine $n_\text{spin} \to \infty$.
Then $\s$ follows horizontal circles around $P$, and the vertical diagonal
cuts join adjacent circles orthogonal to both.

Second, view adjacent bands as deriving from nested cones, as
depicted in Figure~\figref{NestedCones}.
\begin{figure}[htbp]
\centering
\includegraphics[width=0.5\linewidth]{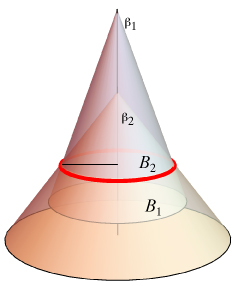}
\caption{Nested cones determined by adjacent bands.}
\figlab{NestedCones}
\end{figure}
Let $\b$ be the half-angle of a cone in $\R^3$, and let $\a$ be the
total surface angle at the cone apex, i.e., the angle of the wedge when
the cone is cut open along a generator and flattened to $\R^2$.
Then $\a = 2 \pi \sin \b$, which implies that these angles grow and shrink
together monotonically.
Let $B_1$ be the lower band and $B_2$ the upper band, sharing a circle
of radius $r$. Then $\b_1 < \b_2$ and so $\a_1 < \a_2$.

Let $\r$ be the planar layout curve with surface to the right,
and $\l$ the curve with surface to the left.
Each unfolded band $B$ is bounded by $\r$ and $\l$, each of
which is an arc of a circle centered on the image of the apex
of the cone containing $B$.

Consider again two adjacent bands $B_1$ and $B_2$ sharing a horizontal circle of $\s$.
$B_1$ is right of $\s$ and $B_2$ left of $\s$ on $P$.
$B_1$ unfolds to a strip bounded by circle arcs of radius $r_1$ strictly larger
than the radius $r_2$.
The shared seam between $B_1$ and $B_2$ therefore unfolds to $\r_1$ and $\l_2$
as illustrated in Figure~\figref{ConeCircsn16k40}.
\begin{figure}[htbp]
\centering
\includegraphics[width=0.75\linewidth]{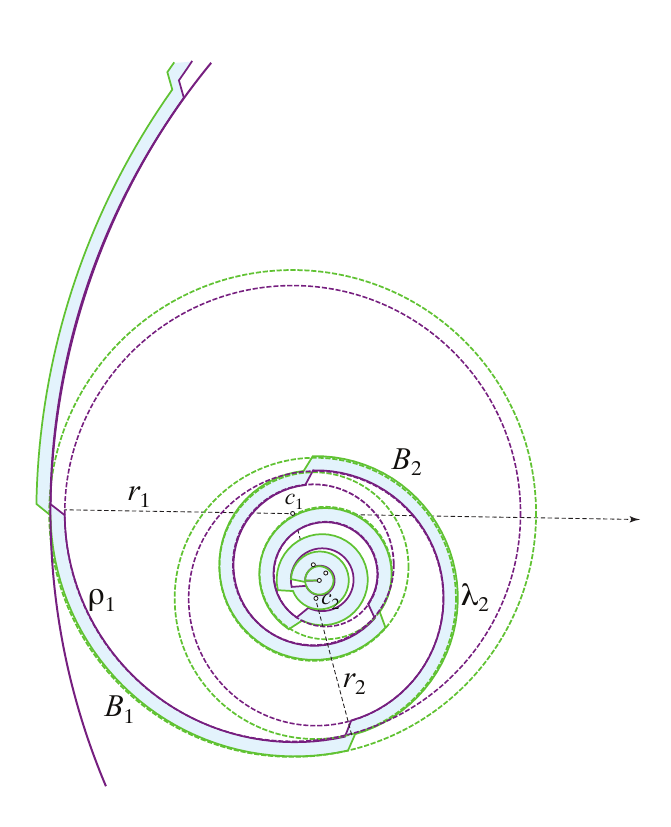}
\caption{Portion of a spiral unfolding with band circles illustrated
(polyhedron of revolution not shown).
We must have $r_1 > r_2$, and $\r_1$ joining smoothly with $\l_2$,
and so the $B_2$ annulus is tangent to and nested inside the $B_1$ annulus.}
\figlab{ConeCircsn16k40}
\end{figure}
The arcs $\r_1$ and $\l_2$ share a common tangent at the orthogonal cuts to adjacent
bands, when $n_\text{spin} \to \infty$.
The centers of the circle arcs are aligned so that $c_1$ and $c_2$ are
collinear with that point of common tangency.

So now it is clear that each band unfolding sits inside an annulus,
with each annulus nested inside and tangent to its $z$-lower mate.
Thus the bands cannot intersect one another except where they join.
Of course the same holds at each ``end" of any polyhedron of revolution $P$,
with the nesting occurring to the other side of the unfolding.

The only possible overlap occurs for small $n_\text{spin}$, when
two regular polygons are separated near the unfolding of the apex of $P$,
as we saw in Figure~\figref{PolRevr4n6ThreeLay}.
But for any apex curvature $\o$, regardless of how small, there is
some $n_\text{spin}$ that avoids overlap, as do all larger $n_\text{spin}$ values.

\subsection{Relationship to arbitrary $P$.}
For arbitrary $P$, the bands are not as circular, and do not necessarily
lie on cones. But one can see an analogous structure
to spiral unfoldings: bands are unfolded and attached,
increasing in ``radii" away from the apexes of $P$.
The more vertices of $P$, and the more circular each cross-section, the
closer will a spiral unfolding resemble an unfolding of a polyhedron
of revolution. This incidentaly reinforces the intuition of why nonoverlap is rare:
even with polyhedra of revolution, the nesting to avoid overlap is delicate.

\subsection{Nonconvex Polyhedra of Revolution}
We end this section 
with a curiosity: some spiral unfoldings of nonconvex polyhedra avoid overlap.
Figure~\figref{Hourglass3figs} shows a nonconvex polyhedron of revolution
and a nonoverlapping unfolding. But this cannot be pushed too far, as the
right overlapping unfolding demonstrates.
\begin{figure}[htbp]
\centering
\includegraphics[width=1.0\linewidth]{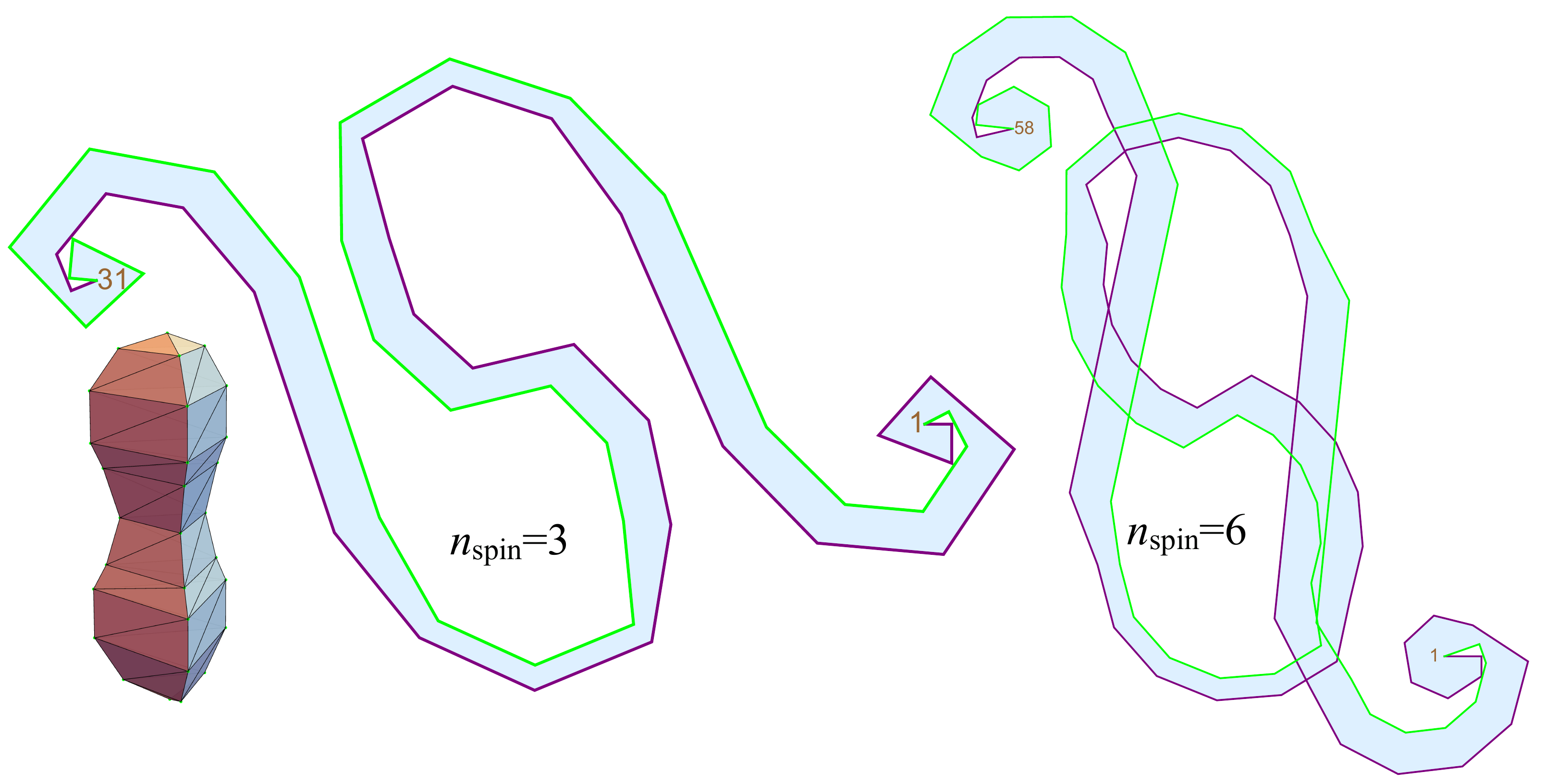}
\caption{A peanut-shaped polyhedron of revolution,
for two different $n_\text{spin}$ values (sharing the same curve $C$).}
\figlab{Hourglass3figs}
\end{figure}

\section{Conjecture}
\seclab{Conjecture}
We mentioned in Section~\secref{Introduction} that it was posed as an
open question in~\cite{lddss-zupc-10} whether or not every convex polyhedron
has a nonoverlapping zipper unfolding.
The spiral unfoldings we explored here are a narrow class of zipper unfoldings,
and shed no direct light on the open problem.
Rather than require a spiral to pass through the vertices in vertical order,
one could imagine loosening the definition to allow spiral-like paths that
rely more on intrinsic surface features rather than on an extrinsic vertical sorting.
I have explored this direction enough to know that some overlaps can be avoided
with more general spirals. Nevertheless, this effort has led me to conjecture
that the answer to the Lubiw et al. open problem is \emph{No}: there are polyhedra
whose every zipper unfolding overlaps.

I reach this conclusion through two hunches.
First, although there are a vast number of possible zipper unfoldings for
polyhedra with many vertices $n$,
if the vertices are sprinkled uniformly but unorganized, spiral-like
paths are the only real options.
Second, for polyhedra that are close to spheres with many unorganized vertices,
any spiral will overlap.
So I suggest two similar candidates for counterexamples:
(1)~$P$ is the convex hull of a large number of random points on a sphere;
(2)~$P$ is geodesic dome, but with the vertices perturbed slightly to break all symmetries.

Figure~\figref{GeoDomeCex6} shows the second example,
\begin{figure}[htbp]
\centering
\includegraphics[width=0.60\linewidth]{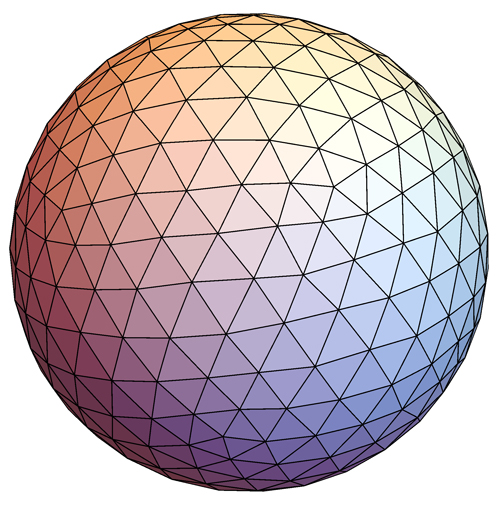}
\caption{Geodesic dome with perturbed vertices.}
\figlab{GeoDomeCex6}
\end{figure}
\begin{figure}[htbp]
\centering
\includegraphics[width=1.00\linewidth]{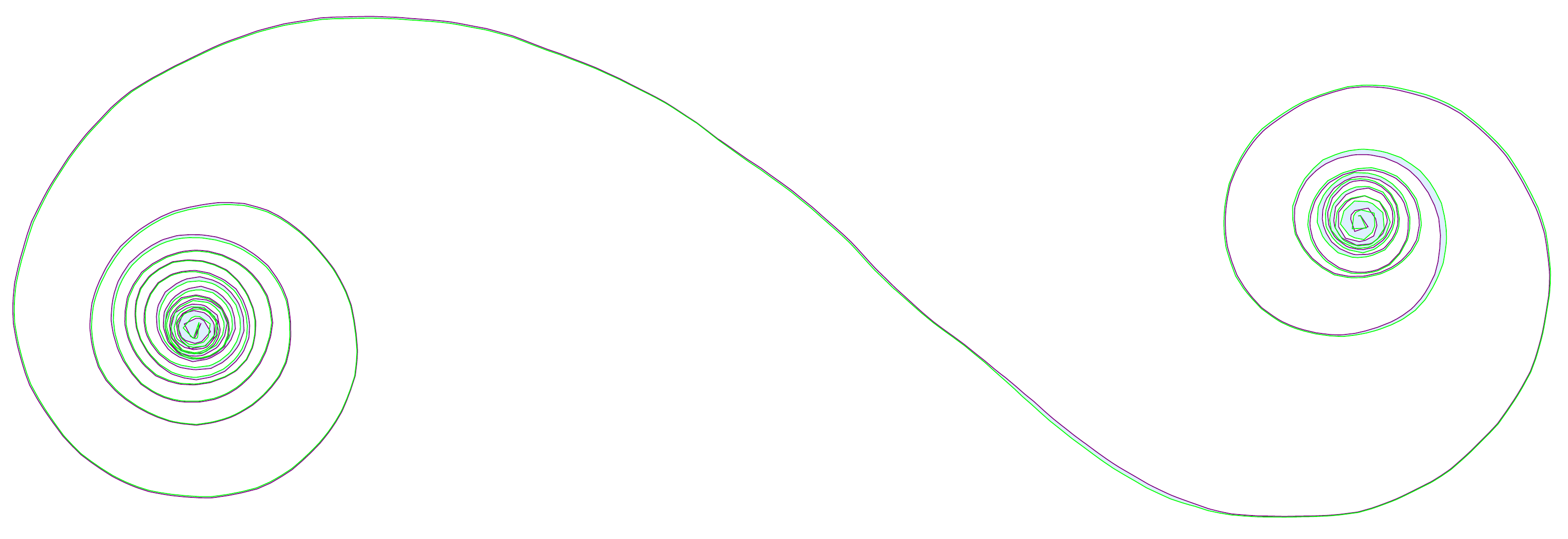}
\caption{Overlapping spiral unfolding of a geodesic dome.}
\figlab{GeoDome3Rand_Lay}
\end{figure}
and Figure~\figref{GeoDome3Rand_Lay} shows a spiral unfolding.
The irregularity of the vertex positions causes significant overlap along
many turns of the spiral.

\bibliographystyle{alpha}
\newcommand{\etalchar}[1]{$^{#1}$}

\end{document}